\documentclass[conf]{new-aiaa}
\usepackage[utf8]{inputenc}

\usepackage{graphicx}
\usepackage{amsmath}
\usepackage[version=4]{mhchem}
\usepackage{siunitx}
\usepackage{longtable,tabularx}

\usepackage{listings}
\usepackage{booktabs}
\usepackage{threeparttable}
\usepackage{caption}
\usepackage{subcaption}

\title{Open-source Framework for Transonic Boundary Layer Natural Transition Analysis over Complex Geometries in Nektar++}

\author{Ganlin Lyu\footnote{PhD Candidate, Department of Aeronautics, g.lyu19@imperial.ac.uk.} }
\affil{Imperial College London, London, United Kingdom, SW7 2AZ}
\author{Chao Chen\footnote{Chief Engineer, Department of Aerodynamics and Aeroacoustics, chenchao@comac.cc.} and Xi Du\footnote{Deputy Manager, Department of Aerodynamics and Aeroacoustics, duxi@comac.cc.}}
\affil{Beijing Aircraft Technology Research Institute of COMAC, Beijing, 102211, China}
\author{Shahid Mughal\footnote{Lecturer, Department of Mathematics, s.mughal@imperial.ac.uk.} and Spencer J. Sherwin\footnote{Professor, Department of Aeronautics, s.sherwin@imperial.ac.uk.}}
\affil{Imperial College London, London, United Kingdom, SW7 2AZ}

\begin{document}

\maketitle

\begin{abstract}
We introduce an open-source and unified framework for transition analysis for laminar boundary layer natural transition at transonic conditions and over complex geometries, where surface irregularities may be present. Different computational tools are integrated in the framework, and therefore overcomes the difficulties of two separate and usually quite disparate processes when using $e^N$ method for transition analysis. To generate a baseflow with desired pressure distribution, appropriate pressure compatible inflow boundary condition needs to be developed and enforced. We first derive the system for 1D numerical stability analysis for boundary conditions, and construct three types of pressure compatible inflow. We demonstrate that the  entropy-pressure compatible inflow is stable unlike other choices. Compared with the steady baseflow computation, the unsteady simulation for the disturbance field is more challenging for compressible flows because of complex wave reflections, which can easily contaminate the results. We therefore introduce the two main sources of wave decontamination and corresponding methods to obtain clean signal. The workflow within the framework is then  verified by computing the disturbance development in 2D flat plate boundary layer flows at Mach $0.8$. The $N$-factors over a clean flat plate and a flat plate with a forward-facing step are generated, and agree well with the results from the reference. Following the verified workflow, We then analyze the disturbance growth on a wing section of the CRM-NLF model. The $N$-factor on a 2D simulation is generated and studied.

\end{abstract}

\section{Nomenclature}

{\renewcommand\arraystretch{1.0}
\noindent\begin{longtable*}{@{}l @{\quad=\quad} l@{}}
$t$ & time \\
$x$, $y$, $z$ & coordinates in the Cartesian coordinate system \\
$Y$ & wall-normal coordinate in the body-fitted coordinate system (distance to the wall) \\
$U$, $W$  & streanwise and spanwise velocity components in the body-fitted coordinate system \\
$T$  & temperature \\
$\mathbf{Q}$ & vector for conservative variable \\
$\mathbf{U}$ & vector for characteristic variable \\
$\mathbf{F}$ & flux \\
$\tilde{\mathbf{f}}$ & numerical flux \\
$\mathbf{n}$ & unit normal vector \\
$\mathbf{C}$ & coefficient matrix \\
$\lambda$ & eigenvalue of a matrix \\
$\Delta$ & filter width for Selective Frequency Damping \\
$\chi$  & control coefficient for Selective Frequency Damping \\
$L$   & length scale of the geometry \\
$u'$  & disturbance on $u$ (the velocity component in the $x$-direction)\\
$f$ & frequency of disturbance \\
$\alpha$ & streamwise wave number \\
$\beta$ & spanwise wave number \\
$C_p$& pressure coefficient \\
\end{longtable*}}

%%%%%%%%%%%%%%%%%%%%%%%%%%%%%%%%%%%%%%%%%%%%%%%%%%%%%%%%%%%%%%%%%%%%%%%%%%%%%%
\section{Introduction}
\label{sec::introduction}
\lettrine{L}{amiar} boundary layer transition and the subsequent turbulent boundary layer for external flows are of particular interest in the aerospace industry since a good aerodynamic design of a vehicle is closely dependent on the correct prediction of transition onset and turbulence features. However, the numerous mechanisms and their complex interactions are still the topic of intensive research in the community. The current work is motivated by designing high fidelity simulation tools to study boundary layer transition at realistic Reynolds numbers and transonic conditions over wings, where the surface may not be smooth but contains surface irregularities.

In aerospace applications of interest, transonic flows are compressible, and external flows experience a freestream with low turbulence intensity and background noise, which enables the transition process to be initiated by the linear growth stage of disturbances. This physical setting makes the $e^N$ method (also known as the $N$-factor method) a common and suitable tool for the prediction of flow transition.

The use of the $e^N$ method involves two major steps: (i) the baseflow computation and (ii) the prediction of the disturbance' growth. Since the disturbance development is sensitive to the baseflow profiles, accurate computation of the baseflow is critical to a successful transition prediction. On clean geometries the baseflow can be computed by a computationally fast and cheap boundary layer solver. Then the disturbance fields are solved according to the Linear Stability Theory (LST) or Parabolized Stability Equations (PSE) \cite{cooke2020modelling}, and the transition prediction analysis can then be undertaken. However, the baseflow over geometries with sufficiently large imperfection such as steps and gaps, cannot typically be computed through a boundary layer equations solver particularly in the presence of local recirculation, i.e. separation bubbles. Therefore, a Navier-Stokes  solver needs to be used to obtain the baseflow. As for the disturbance fields, a linearized solver or full Navier-Stokes solver has to then be applied, since typically PSE models fail to correctly capture any rapid short scale variations in the baseflow; as well as some uncertainty of correctness of PSE modelling when dealing with large segments of locally confined reversed flow.

When different computational tools are involved in these two main steps, data conversion issues can arise, typically involving difficulties interpolating and rescaling to  alternative grids as accurately as possible. This causes losses in precision of data and extra workload for researchers and engineers. A particular example would be that of an abrupt change in the vicinity of a localized stepped feature -- interpolation using a standard cubic spline of the boundary layers for the consequent instability analysis may give rise to overshoots and unphysical flow gradients, which then impact the instability analysis. A set of integrated tools is therefore desirable. However, to the best of authors' knowledge although different research groups have developed their own tools for boundary layer analysis and flow transition prediction, relatively few, if any, are available as open-source to the wider community. Since developing each of the tools requires specific yet different knowledge, new researchers in the field find it extremely difficult, due to unavailability of trusted and well documented software, to start their investigation. Therefore, we have developed an open-source and unified framework which overcomes the difficulties of the two-stage process (i.e two separate and usually quite disparate baseflow and linear stability computations, and usually different numerical discretization strategies used in the base flow and instability tools). Our paper will describe a unified approach to modelling flow instabilities, which utilizes the spectral/hp element method framework Nektar++ \cite{cantwell2015nektar++,moxey2020nektar++}, which is coded in C++ and has the merits of being cross-platform and open-source. The high-order solvers in Nektar++ enable accurate boundary layer profile computations over complex geometries (which may well include locally reversed flows), as well as the ability to capture disturbance development with high accuracy.

This paper is organized as follows. Section \ref{sec::workflow} introduces workflow transition analysis. In the baseflow and disturbance fields computation of the workflow, pressure compatibility with the background results is desired. In section \ref{sec::pressure_compatible_inflow} we analyze one-dimensional (1D) stability for inflow boundary condition to achieve pressure compatibility. An entropy-pressure compatible inflow is selected. Section \ref{sec::de-contamination} discusses methods to reduce wave contamination for the disturbance field computation. The ability to accurately capture disturbance development is demonstrated in section \ref{sec::precision}. Section \ref{sec::flatPlate} verifies the workflow using a two-dimensional (2D) transonic flat plate problem with both clean and stepped geometries. Finally, a 2D transitional study on a wing section of CRM-NLF model is provided in section \ref{sec::CRM}. For the better demonstration purpose, some intermediate results from the flat plate case and the CRM case are used in early sections although they are more completely discussed in the last section.

%%%%%%%%%%%%%%%%%%%%%%%%%%%%%%%%%%%%%%%%%%%%%%%%%%%%%%%%%%%%%%%%%%%%%%%%%%%%%%%%%%%%%%%%%%%%
\section{Workflow for transition prediction}
\label{sec::workflow}
%flowchart, airfoil flowchart, LST in NekPy interface, 

\begin{figure}[hbt!]
  \centering
  \includegraphics[width=12cm]{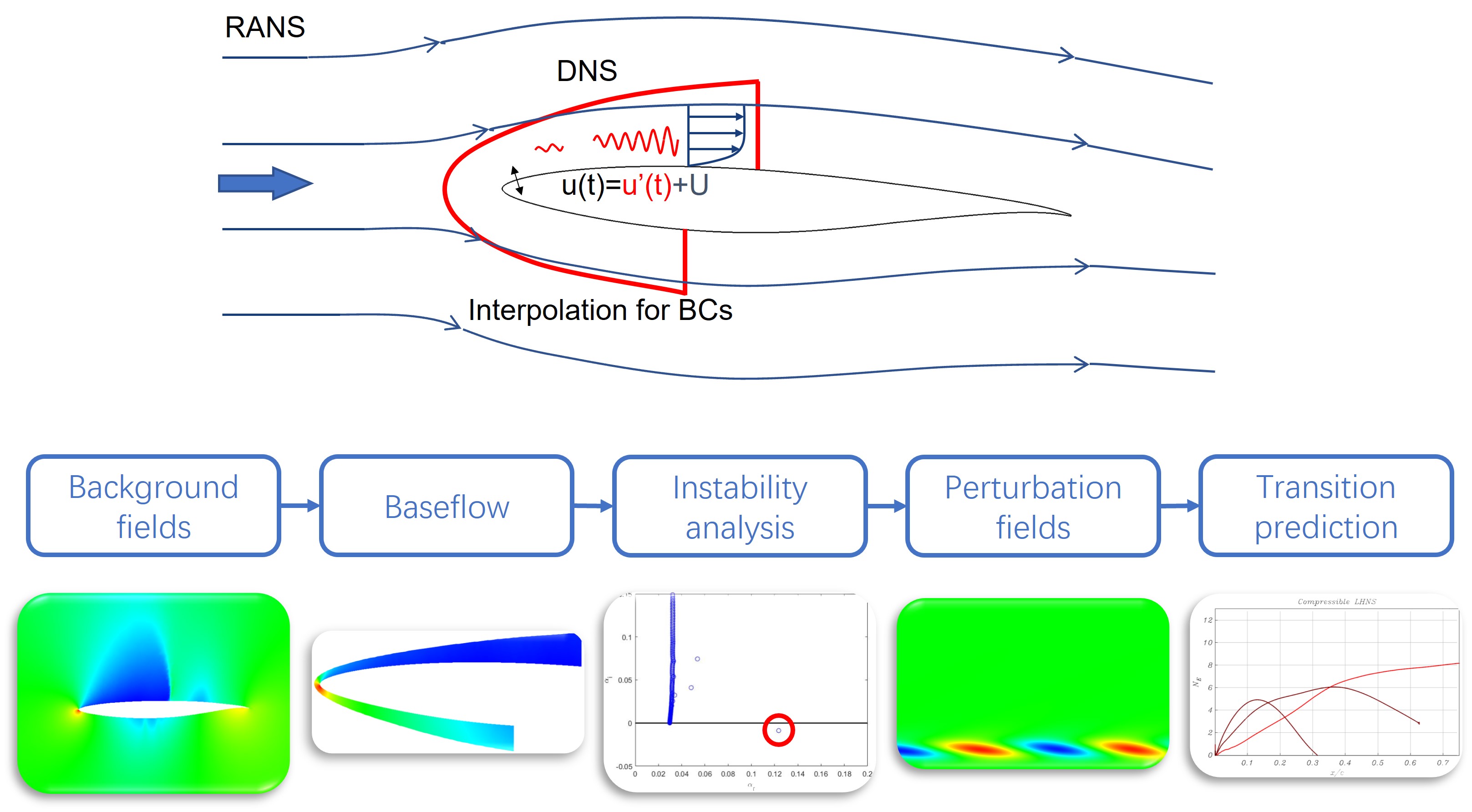}
  \caption{Workflow for transonic boundary layer natural transition analysis.}
  \label{fig_workflow}
\end{figure}

The workflow for transition prediction over a wing section is given in Fig. \ref{fig_workflow}. To reduce the computational cost of high-fidelity simulation of boundary layer flows, we wish to use a near-body, reduced domain, whose outer boundary conditions are interpolated from a computationally cheaper three-dimensional (3D) Reynolds-Averaged Navier-Stokes (RANS) simulation. Taking advantage of its lower cost, the RANS simulation can be carried out over the full geometry such as a wing-fuselage configuration, and thus the three-dimensionality effects and the influence by the fuselage on the field distributions are automatically taken into account.

As a next step we compute the baseflow in the reduced domain, including the near wall laminar boundary layer. The matching of the outer RANS solution with the inner baseflow requires careful treatment of the inflow conditions. For a subsonic inflow condition in a Discontinuous Galerkin (DG) and Riemann-based solver, two conditions can be imposed in the normal direction to the boundary. As the standard condition in a DG solver, the incoming Riemann invariant and the measure of entropy are set, leading to a non-reflecting boundary. However, this does not adequately enforce a compatible pressure condition with the outer RANS simulation. This incompatibility in pressure distribution is undesirable since the pressure load is usually well captured by the Euler or RANS simulation (at least for lift prediction) and since the pressure distribution does not typically vary much over the boundary layer. Moreover, the disturbance development inside the boundary layer is significantly influenced by the pressure distribution. The above reasons therefore make pressure distribution a quantity of interest from the lower fidelity models and is an important property to maintain in the reduced domain, for the higher fidelity simulations that we hope to achieve with Nektar++. To best enforce the pressure compatibility, in this work 
three types of pressure compatible inflow boundary conditions are considered, however the entropy-pressure compatible inflow is adopted since it is stable unlike other choices according to 1D analysis. 
%The entropy is chosen as another condition because one-dimensional stability analysis shows that the incompatibility in inflow entropy can easily lead to an unstable system. 

Before we solve the disturbance fields, sectional instability analysis is carried out along the boundary layer to ascertain and map out the instability parameter space, in terms of frequencies and spanwise wavenumbers of unstable disturbances, likely playing the most dominant role(s) in the transition process in the boundary layer. This is achieved through a local LST spatial instability analysis of boundary layer profiles at different streamwise positions. In this step accurate boundary layer profiles can be extracted interactively using the Nektar++ Python interface (NekPy), where the spatial analysis is undertaken using the LST module, using the open-source part of the CoPSE3d code \cite{mughal1998active,mughal2006stability}. In the LST module the compressible LST equations are discretized with 4th-order finite differences. A generalized QZ method is used for eigenvalues and the discrete mode selected. This is then further refined using inverse Rayleigh iteration to get a highly accurate grid independent eigenvalue and eigenfunction.

Next, having identified the most dominant disturbances’ frequencies and spanwise wavenumbers, the prescribed disturbances are then introduced in the reduced domain. To excite the disturbances, artificial receptivity is adopted by setting part of the wall near the leading edge as a suction-blowing interface, operating at the selected frequencies. Finally, the envelope ($N$-factor) of growth curves ($n$-factors) can be generated, and the transition onset is considered at the position where the envelope exceeds a user prescribed threshold.

It is worth mentioning although the current work focuses on compressible flows, the aforementioned approach is also equivalently applicable to laminar boundary layer natural transition predictions for incompressible flows. The differences lie in the choices of solvers in Nektar++ to compute the baseflows and disturbance fields, as well as the necessity to reduce signal contamination by the wave reflections.

%%%%%%%%%%%%%%%%%%%%%%%%%%%%%%%%%%%%%%%%%%%%%%%%%%%%%%%%%%%%%%%%%%%%%%%
\section{Stability analysis for pressure compatible inflow boundary condition}
\label{sec::pressure_compatible_inflow}

Having obtained an outer RANS solution, the desired, pressure compatible baseflow is computed through appropriate boundary condition enforcement. For a reduced domain over wing-shaped geometries, the vast of the domain is rounded by the inflow boundary. A pressure compatible inflow therefore needs to be constructed. In a DG based compressible flow simulation, two or three conditions can be imposed in the normal direction to the inflow boundary depending on whether the flow is subsonic or supersonic in the normal direction. If it is supersonic, the pressure compatibility is automatically guaranteed as all information comes from the upstream information. However, at the subsonic condition one piece of information is determined from the interior domain data. The first step to construct a pressure compatible inflow is to select the other desired compatible conditions in addition to the pressure. However, not all the selections leads to stable numerical solution or even a well-posed problem. We therefore analyze the 1D stability of three possible pressure compatible inflows. To set up the system for stability analysis, we consider a piece-wise constant approximation of 1D Euler equations which is then linearized in an upstream element with inflow boundary \cite{lyu2022wellposed}.

%In the following sub-sections, the piece-wise constant approximation of 1D Euler equations are first linearized in an upstream element with inflow boundary, and the linear systems for the inflow boundary conditions to be stable are derived. An entropy-pressure compatible inflow boundary condition is selected from the full range of possibilities.

\subsection{Linearized DG approximation to the 1D Euler equations}

The 1D Euler equations take the form
\begin{equation}
\label{eq::Euler1D_differenial}
  \frac{\partial \mathbf{Q}}{\partial t}
+ \frac{\partial \mathbf{F(\mathbf{Q})}}{\partial x} = \mathbf{0}
\end{equation}
where $\mathbf{Q}$ is the state vector of conservative variables, $\mathbf{F}$ is vector for inviscid flux
\begin{equation}
  \mathbf{Q} =\left(
    \begin{array}{c c c}
      \rho, \rho u, E
    \end{array} \right)^T, \hspace{0.5 cm}
  \mathbf{F} =\left(
    \begin{array}{c c c}
      \rho u, p + \rho u^2, u (E + p)
    \end{array} \right)^T
\end{equation}

In the DG based spectral/hp element method, the approximation in the elements are independent except for the weak coupling through numerical flux at the shared boundaries by adjacent elements. The numerical flux is therefore the approach for element-wise boundary condition. As mentioned earlier, to analyze the inflow conditions we focus on a single element located at the very upstream position of the domain. In such a way the element would have an inflow boundary and one internal boundaries. To approximate the solution in the element, we first multiply Eq. \ref{eq::Euler1D_differenial} by a test function $\phi$ and then integrate in the domain $\Omega$ by part, which gives
\begin{equation}
\label{eq::Euler1D_DG_1}
  \int_{\Omega} \frac{\partial\mathbf{Q}}{\partial t} \phi dx  + 
  \phi \mathbf{F}\cdot \mathbf{n} |_{\partial \Omega} - \int_{\Omega} \nabla \phi \cdot \mathbf{F} = 0
\end{equation} 
For simplicity we consider the piece constant approximation
\begin{equation}
\label{eq::pieceConstantApproximation}
  \mathbf{Q} \simeq \hat{\mathbf{Q}}(t) \phi(x), \hspace{0.5 cm}
  \phi(x) = \left\{ 
  \begin{aligned} 
    & 1, \quad x \in \Omega  \\
    & 0, \quad x \notin \Omega 
  \end{aligned} \right.
\end{equation} 
By replacing the boundary flux $\mathbf{F}$ by the numerical flux $\tilde{\mathbf{f}}$, Eq. \ref{eq::Euler1D_DG_1} becomes
\begin{equation}
\label{eq::Euler1D_DG_2}
  \frac{d\hat{\mathbf{Q}}}{d t} \Delta x = \tilde{\mathbf{f}}^{w} - \tilde{\mathbf{f}}^{e} 
\end{equation}
where the flow direction is from west to east, $\Delta x = x^e-x^w$ is the length of the 1D element and the superscript $w$ and $e$ denote the west and east boundaries. The numerical flux is typically computed through a Riemann solver, and is the function of the internal state and external state. For the numerical flux at the solution domain boundary, where there is no adjacent element, a ghost state  $\hat{\mathbf{Q}}_{gh}$ is introduced instead of the external state so that we have
\begin{equation}
  \tilde{\mathbf{f}} = \tilde{\mathbf{f}}(\hat{\mathbf{Q}}_{gh},\hat{\mathbf{Q}}_{int})
\end{equation}
In the standard use of a Riemann boundary condition the external or ghost state is independent of the internal state and the solution of the Riemann solver will provide a known boundary state $\hat{\mathbf{Q}}_b$, i.e.
\[
\hat{\mathbf{Q}}_b(\hat{\mathbf{Q}}_{gh},\hat{\mathbf{Q}}_{int}). \]
In this circumstances $\hat{\mathbf{Q}}_b$ is not known explicitly but rather inferred through  $\hat{\mathbf{Q}}_{gh}$ and  $\hat{\mathbf{Q}}_{int}$ and the Riemann solution procedure \cite{mengaldo2014guide}. Nevertheless it is also possible to explicitly impose as many conditions as there are incoming characteristics on the domain boundary,  for example a prescribed freestream quantity such as pressure or a zero normal velocity at an inviscid wall. For this situation we  introduce 
 $\hat{\mathbf{Q}}_{ref}$ to denote an external reference state which when provided as input (or the ghost state) to the Riemann solver in combination with the internal state it enforces the desired condition on $\hat{\mathbf{Q}}_b$ and subsequently $\tilde{\mathbf{f}}(\hat{\mathbf{Q}}_b)$. However in this case the reference state $\hat{\mathbf{Q}}_{ref}$ is now dependent on $\hat{\mathbf{Q}}_{int}$ and the desired conditions we wish to impose such that
\begin{equation}
\begin{aligned}
 \tilde{\mathbf{f}} & = \tilde{\mathbf{f}}(\hat{\mathbf{Q}}_{ref}(\hat{\mathbf{Q}}_{int}), \hat{\mathbf{Q}}_{int}) \\ 
  & = \tilde{\mathbf{f}}(\hat{\mathbf{Q}}_{ref},\hat{\mathbf{Q}}_{int}) 
\end{aligned}
\end{equation}

The linearized form of Eq. (\ref{eq::Euler1D_DG_2}) is then given by
\begin{equation}
\label{eq::LinStability1D_Q}
  \frac{d \left(\delta \hat{\mathbf{Q}}_{int} \right)}{d t} \Delta x = \left( 
  \frac{\partial \tilde{\mathbf{f}}^w}{\partial \hat{\mathbf{Q}}_{int}} - \frac{\partial \tilde{\mathbf{f}}^e}{\partial \hat{\mathbf{Q}}_{int}} \right)  \delta \hat{\mathbf{Q}}_{int} + \left( 
  \frac{\partial \tilde{\mathbf{f}}^w}{\partial \hat{\mathbf{Q}}_{ref}^w}  - \frac{\partial \tilde{\mathbf{f}}^e}{\partial \hat{\mathbf{Q}}_{ref}^e} \right) \delta \hat{\mathbf{Q}}_{ref}
\end{equation}
where $\delta \hat{\mathbf{Q}}_{int}$ is the time-dependent disturbance on the internal state and $\delta \hat{\mathbf{Q}}_{ref}$ is the constant disturbance on the reference state. (The disturbance on the west boundary is assumed equals to that on the east boundary.)

In 1D analysis it is convenient to transform the linearized form in Eq. \ref{eq::LinStability1D_Q} to use the characteristic variables as independent variables 
\begin{equation}
\label{eq::characteristicVars}
  \mathbf{U} =\left(
  \begin{array}{c c c}
    R^+, R^0,  R^-
  \end{array} \right)^T = \left(
  \begin{array}{c c c}
    u+\frac{2}{\gamma-1} c,   S,  u-\frac{2}{\gamma-1} c
  \end{array} \right)^T
\end{equation}
where $c=\sqrt{\gamma p/\rho}$ is the speed of sound, $S = \frac{p}{\rho^\gamma}$ is the measure of entropy, and $\gamma$ is the specific heat ratio.

By assuming the baseflow is steady and therefore uniform because of one-dimensionality, the 1D linearized system in characteristic variables takes the form
\begin{equation}
\label{eq::LinStability1D_U}
  \frac{d}{d t} \left(\delta \hat{\mathbf{U}}_{int} \right) =
  \frac{1}{\Delta x} \mathbf{C}_{int} \delta \hat{\mathbf{U}}_{int} + \frac{1}{\Delta x} \mathbf{C}_{ref} \delta \hat{\mathbf{U}}_{ref} 
\end{equation}
where the coefficient matrices are defined as 
\begin{equation}
\label{eq::CoefMatrix_int}
  \mathbf{C}_{int} = 
  \left(\frac{\partial \hat{\mathbf{Q}}}{\partial \hat{\mathbf{U}}} \right)^{-1}
  \frac{\partial \tilde{\mathbf{f}}}{\partial \hat{\mathbf{Q}}}
  \frac{\partial \hat{\mathbf{Q}}}{\partial \hat{\mathbf{U}}}
  \left(
  \frac{\partial \hat{\mathbf{U}}_b^w}{\partial \hat{\mathbf{U}}_{int}} - \frac{\partial \hat{\mathbf{U}}_b^e}{\partial \hat{\mathbf{U}}_{int}}
  \right) = \mathbf{C}_{1}\mathbf{C}_{2,int},
\end{equation}
\begin{equation}
\label{eq::CoefMatrix_ref}
  \mathbf{C}_{ref} = 
  \left(\frac{\partial \hat{\mathbf{Q}}}{\partial \hat{\mathbf{U}}} \right)^{-1}
  \frac{\partial \tilde{\mathbf{f}}}{\partial \hat{\mathbf{Q}}}
  \frac{\partial \hat{\mathbf{Q}}}{\partial \hat{\mathbf{U}}}
  \left(
  \frac{\partial \hat{\mathbf{U}}_b^w}{\partial \hat{\mathbf{U}}_{ref}^w} - \frac{\partial \hat{\mathbf{U}}_b^e}{\partial \hat{\mathbf{U}}_{ref}^e} \right) = \mathbf{C}_{1}\mathbf{C}_{2,ref}.
\end{equation}

In the above, $\mathbf{C}_{1}$  relates to the uniform baseflow quantities and can be further evaluated as: 
\begin{equation}
\label{eq::CoefMatrix_1}
  \mathbf{C}_1 = \left(\frac{\partial \hat{\mathbf{Q}}}{\partial \hat{\mathbf{U}}} \right)^{-1} \frac{\partial \tilde{\mathbf{f}}}{\partial \hat{\mathbf{Q}}}
  \frac{\partial \hat{\mathbf{Q}}}{\partial \hat{\mathbf{U}}} =\left(
  \begin{array}{c c c}
    c+u & -\frac{c^2}{\gamma(\gamma-1)S} & 0 \\
      0 &         u                      & 0 \\
      0 & -\frac{c^2}{\gamma(\gamma-1)S} & u-c
  \end{array} \right)
\end{equation}
which is similar to the Jacobian matrix $\partial \tilde{\mathbf{f}}/\partial \hat{\mathbf{Q}}$, and therefore the eigenvalues are the well-known $u+c$, $u$, and $u-c$. The remaining  coefficient matrices are
\begin{equation}
\label{eq::CoefMatrix_2_int}
  \mathbf{C}_{2,int} = 
  \frac{\partial \hat{\mathbf{U}}_b^w}{\partial \hat{\mathbf{U}}_{int}} -
  \frac{\partial \hat{\mathbf{U}}_b^e}{\partial \hat{\mathbf{U}}_{int}}
\end{equation}
\begin{equation}
\label{eq::CoefMatrix_2_ref}
  \mathbf{C}_{2,ref} = 
  \frac{\partial \hat{\mathbf{U}}_b^w}{\partial \hat{\mathbf{U}}_{ref}^w} -
  \frac{\partial \hat{\mathbf{U}}_b^e}{\partial \hat{\mathbf{U}}_{ref}^e}
\end{equation}
which contain the normalized coefficients about how the boundary condition will respond to the disturbance of the internal and reference states, respectively.

For the coefficient matrix $\mathbf{C}_{int}$, we assume it can be diagonalised so that 
$$ \mathbf{C}_{int} = \mathbf{P}^{-1} \mathbf{\Lambda} \mathbf{P} $$
where $\mathbf{\Lambda} = diag\left( \lambda_i \right)$ ($i=1,2,3$), and $\lambda_i$ is its eigenvalue. If these eigenvalues are non-zero, the solution to Eq. (\ref{eq::LinStability1D_U}) can be written as
\begin{equation}
\label{eq::LinStability1D_U_solution}
  \mathbf{P} \delta\hat{\mathbf{U}}_{int} (t) = e^{\mathbf{\Lambda} t/\Delta x} \mathbf{P} \delta\hat{\mathbf{U}}_{int} (0) + \left(\mathbf{I} - e^{\mathbf{\Lambda} t/\Delta x}\right) \left( -\mathbf{P} \mathbf{C}_{int}^{-1} \mathbf{C}_{ref} \right)  \delta\hat{\mathbf{U}}_{ref}.
\end{equation}
For a stable solution to a well-posed problem, the initial disturbance on the internal state $\delta\hat{\mathbf{U}}_{int} (0)$ in Eq. (\ref{eq::LinStability1D_U_solution}) should decay. Physically we understand that the disturbance waves will leave the domain through the boundaries without generating a stronger reflection, meanwhile the disturbance on the reference state enters the domain. The solution should exponentially converge to the reference value of the boundary conditions. This result is achieved if, and only if, the following conditions are satisfied
\begin{equation}
\label{eq::LinStability1D_U_requirement}
  \left\{
  \begin{aligned}
  &  Re(\lambda_i) < 0, \hspace{0.3 cm} i=1,2,3\\
  &  \mathbf{C}_{ref} = - \mathbf{C}_{int}.
  \end{aligned}
  \right.
\end{equation}
where the second condition is equivalent to a opposite relation for the second part of the coefficient matrices
\begin{equation}
\label{eq::LinStability1D_U_requirement_equivalent}
  \mathbf{C}_{2,ref} = - \mathbf{C}_{2,int}
\end{equation}
For a boundary state provided by a Riemann solver, however, the characteristic must equal to that from either internal state or the reference state, which finally guarantee the satisfaction of Eq. \ref{eq::LinStability1D_U_requirement_equivalent}. Therefore the stability of the linearized system can be directly analyzed through the eigenvalue of $\mathbf{C}_{int}$, whose detailed form depends on the construction of boundary conditions.

\subsection{Construction and stability analysis for pressure compatible Riemann inflows}

\begin{figure}[hbt!]
  \centering
  \includegraphics[width=6cm]{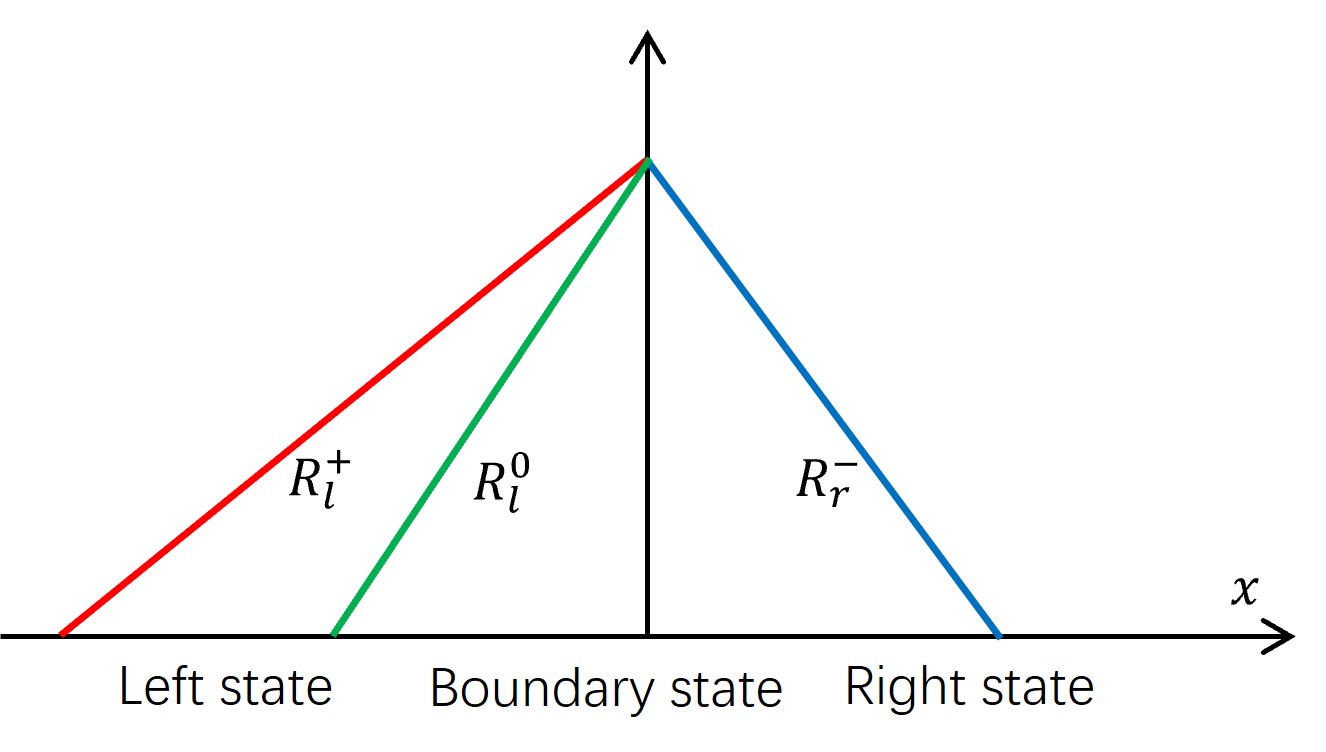}
  \caption{Characteristic lines for a subsonic boundary.}
  \label{fig_characteristics}
\end{figure}

The construction of a boundary condition is derived from the expression of the boundary state, provided by the Riemann solver using the characteristics on both sides of the interface, as is shown in Fig. \ref{fig_characteristics}. For a subsonic inflow, the boundary state is computed based on the two specified quantities together with the outward-propagating characteristic from the internal state ($R_r^-$ in Fig. \ref{fig_characteristics}).

%For a subsonic inflow, the Riemann solver implicitly selects the outward-propagating characteristic from the internal state and the two inward-propagating characteristics from the ghost state. 

To achieve pressure compatibility, we construct three possible subsonic inflows: 
\begin{itemize}
  \item Entropy-pressure compatible inflow (SP);
  \item Velocity-pressure compatible inflow (UP);
  \item Momentum-pressure compatible inflow (MP).
\end{itemize}
whose boundary states are listed below in characteristic form 

\begin{equation}
  \hat{\mathbf{U}}_{b}^{SP} =\left(
    \begin{array}{c c c}
      R_{b}^+,  R_{b}^0,  R_{b}^-
    \end{array} \right)^T = \left(
    \begin{array}{c c c}
      \left(R_{ref}^+ - R_{ref}^- + R_{int}^-\right),  R_{ref}^0,  R_{int}^-
    \end{array} \right)^T
    \label{eq::SP_I_bnd}
\end{equation}

\begin{equation}
  \hat{\mathbf{U}}_{b}^{UP} = \left(
    \begin{array}{c c c}
      \left(R_{ref}^+ + R_{ref}^- - R_{int}^- \right),    
      \left[ \frac{ \left( R_{ref}^+ + R_{ref}^- \right) -2 R_{int}^-}{ R_{ref}^+ - R_{ref}^-} \right]^{2\gamma} R_{ref}^0,  R_{int}^-
    \end{array} \right)^T
    \label{eq::UP_I_bnd}
\end{equation}

\begin{equation}
  \hat{\mathbf{U}}_{b}^{MP} = \left(
    \begin{array}{c c c}
      \left[\frac{\eta^2}{2} \left(R_{ref}^+ + R_{ref}^- \right) + \frac{\eta}{2}\left(R_{ref}^+ - R_{ref}^-\right) \right],   
      \eta^{2\gamma} R_{ref}^0, R_{int}^-
    \end{array} \right)^T
    \label{eq::MP_I_bnd}
\end{equation}
where
\[
\label{eq::MP_BC_3}
  \eta = \sqrt{\frac{u_{b}}{u_{ref}}} = \frac{1}{\gamma-1} \frac{c_{ref}}{u_{ref}} - \sqrt{\left( \frac{1}{\gamma-1} \frac{c_{ref}}{u_{ref}} \right)^2 + \frac{R_{int}^-}{u_{ref}}  }\]
To complete the stability analysis, the boundary condition at the outflow also needs to be provided. In what follows we will enforce the invariant compatible condition, which is satisfied by default at the internal outflow boundary. Using the boundary state expressions in Eqs. \ref{eq::SP_I_bnd} -- \ref{eq::MP_I_bnd}, we derive the

\begin{equation}
\left\{
\begin{aligned}
  & \mathbf{C}_{int}^{SP} = \mathbf{C}_{1} \left(
  \begin{array}{c c c}
    -1 &  0 & 1 \\
     0 & -1 & 0 \\
     0 &  0 & 1
 \end{array} \right) \\
  & \lambda_1^{SP} = -(c+u), \hspace{0.5 cm}
  \lambda_2^{SP} = -u, \hspace{0.5 cm}
  \lambda_3^{SP} = -(c-u)
\end{aligned}
\right.
\end{equation}

\begin{equation}
\left\{
\begin{aligned}
  & \mathbf{C}_{int}^{UP} = \mathbf{C}_{1} \left(
  \begin{array}{c c c}
    -1 &  0 & -1 \\
     0 & -1 & -(\gamma-1)\frac{c}{\rho^{\gamma-1}} \\
     0 &  0 & 1
 \end{array} \right) \\
  & \lambda_1^{UP} = -(c+u), \hspace{0.5 cm}
  \lambda_{2,3}^{UP} = \pm \sqrt{-u(c-u)}
\end{aligned}
\right.
\end{equation}

\begin{equation}
\left\{
\begin{aligned}
  & \mathbf{C}_{int}^{MP} = \mathbf{C}_{1} \left(
  \begin{array}{c c c}
    -1 &  0 & -2 \frac{c+(\gamma-1)u}{c-(\gamma-1)u} \\
     0 & -1 & -2 (\gamma-1) \frac{c^2}{\rho^{\gamma-1} \left[c-(\gamma-1)u\right]} \\
     0 &  0 & 1
 \end{array} \right) \\
  & \lambda_1^{MP} = -(c+u), \hspace{0.5 cm}
  \lambda_{2,3}^{MP} = \frac{\pm \sqrt{\sigma} - cu + c^2 + \gamma c u}{2(c+u-\gamma u)}
\end{aligned}
\right.
\end{equation}
where
\[
\sigma = c^4 + 2(\gamma-3) c^3 u + \gamma(\gamma+6) c^2 u^2 - 3c^2 u^2 + 4(1-\gamma^2) c u^3 + 4(\gamma -1)^2 u^4 \]

\begin{figure}[hbt!]
  \centering
  \includegraphics[width=0.49\textwidth]{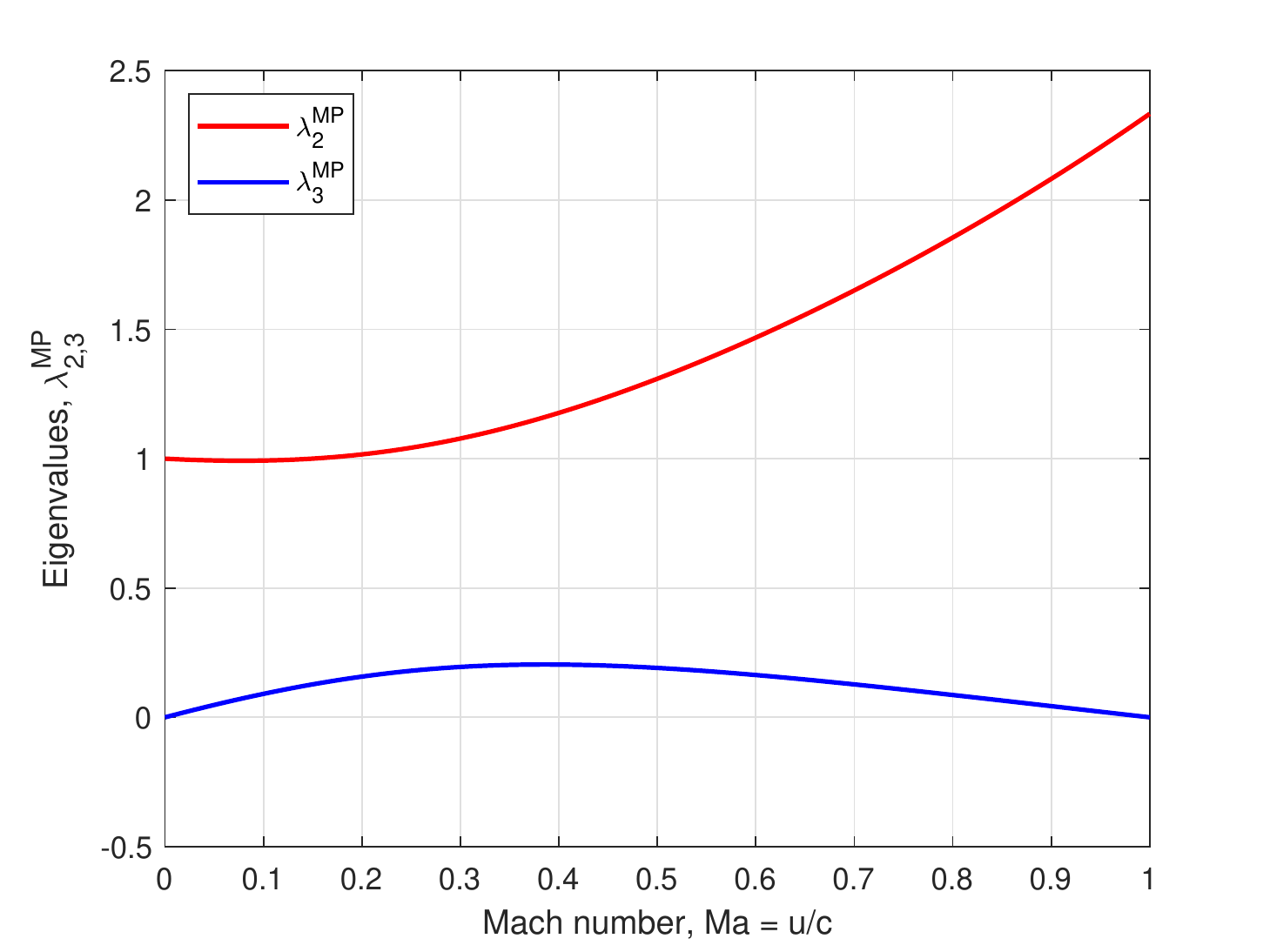}
  \caption{Variation of $\lambda_{2,3}^{MP}$ with Mach number in subsonic region.}
  \label{fig_eigenvalues_MP}
\end{figure}

In the above, the entropy-pressure compatible inflow is stable since $\lambda_{1,2,3}^{SP}$ are negative for subsonic flows. The purely imaginary nature of $\lambda_{2,3}^{UP}$ indicates that some components of the internal disturbance will keep oscillating and therefore the velocity-pressure compatible inflow is not desirable for a steady state solution. As for the momentum-pressure compatible inflow, the variation of $\lambda_{2,3}^{MP}$ with Mach number are depicted in Fig. \ref{fig_eigenvalues_MP}, where the eigenvalues are always positive in subsonic region, showing the instability. The stability analysis results for the three pressure compatible inflow are summarized in Table \ref{tab::stability_for_inflow_BCs}, where the only stable candidate is the entropy-pressure compatible inflow. 

It is worth noting that that since the entropy is a function of pressure and density, the the entropy-pressure compatible inflow also guarantee the compatibilities for density and consequently the speed of sound. In multi-dimensional simulations, the tangential velocity can be directly specified as the RANS data, this will also lead to tangential momentum compatibility. More details on the derivation are also provided in Ref. \cite{lyu2022wellposed}.
%{\bf Cite JCP paper in preparation}

\begin{table}[hbt!]
  \centering
  \caption{1D Stability for inflow boundary conditions}
    \begin{tabular}{c c c c c c}
    \toprule
        & Inflow compatibility & Outflow compatibility & Stability \\  
    \midrule
      1 & Entropy-pressure       & Invariant  & Stable  \\
      2 & Velocity-pressure      & Invariant  & Neutral stable \\
      3 & Momentum-pressure      & Invariant  & Unstable \\
    \bottomrule
    \end{tabular}
    \label{tab::stability_for_inflow_BCs}
\end{table}

%%%%%%%%%%%%%%%%%%%%%%%%%%%%%%%%%%%%%%%%%%%%%%%%%%%%%%%%%%%%%%%%%%%%%%%
\section{Methods to reduce wave contamination}
\label{sec::de-contamination}
%sponge, One-dimensional data filtering (Chebyshev polynomials + modified Least squares approximation)

\begin{figure}[hbt!]
  \centering
  \includegraphics[width=11cm]{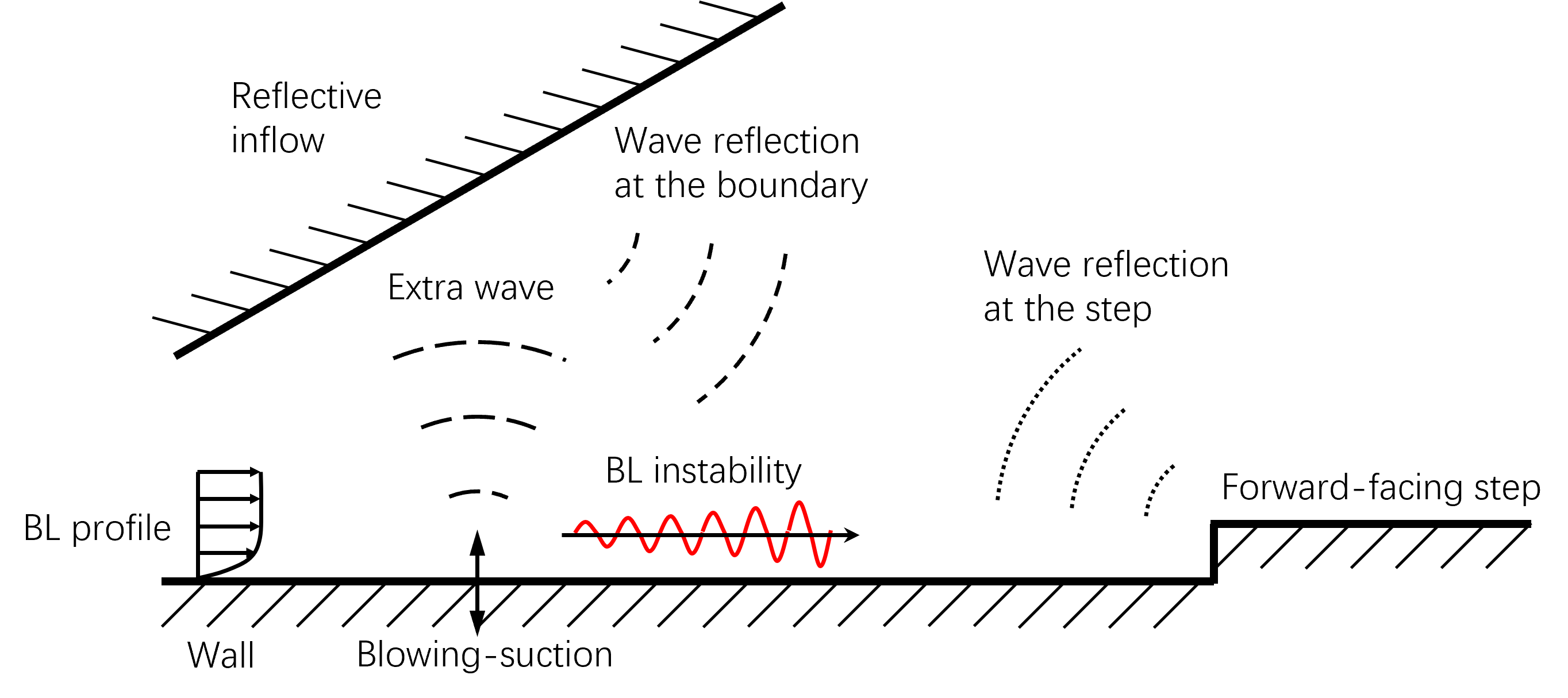}
  \caption{Sketch of the wave reflection at a inflow boundary and at a forward-facing step. The long dashed line represents the extra wave caused by the blowing-suction, the short dashed line represents the reflection wave at the inflow boundary, and the dotted line represents the reflection wave at the step.}
  \label{fig_workflow_waveReflection_sketch}
\end{figure}

Unlike incompressible flows where the speed of sound is assumed infinite so that the information of pressure and velocity influences the whole field instantaneously, the compressible flows update the fields through complex wave propagation due to its mixed parabolic-hyperbolic nature. The emergence of undesired waves may contaminate the disturbance signals of interest, such as the fields for Tollmien–Schlichting (TS) waves and crossflow waves. For the transonic boundary layer natural transition analysis over complex geometries, two main causes of wave contamination are:
\begin{itemize}
  \item Waves generated by blowing-suction and reflected at the inflow boundaries;
  \item Reflected waves at the surface irregularities.
\end{itemize}
both of which are related to the wave reflection at the boundaries of the domain, and a sketch of the reflections are shown in Fig. \ref{fig_workflow_waveReflection_sketch}. 

\begin{figure}[hbt!]
  \centering
  \includegraphics[width=10cm]{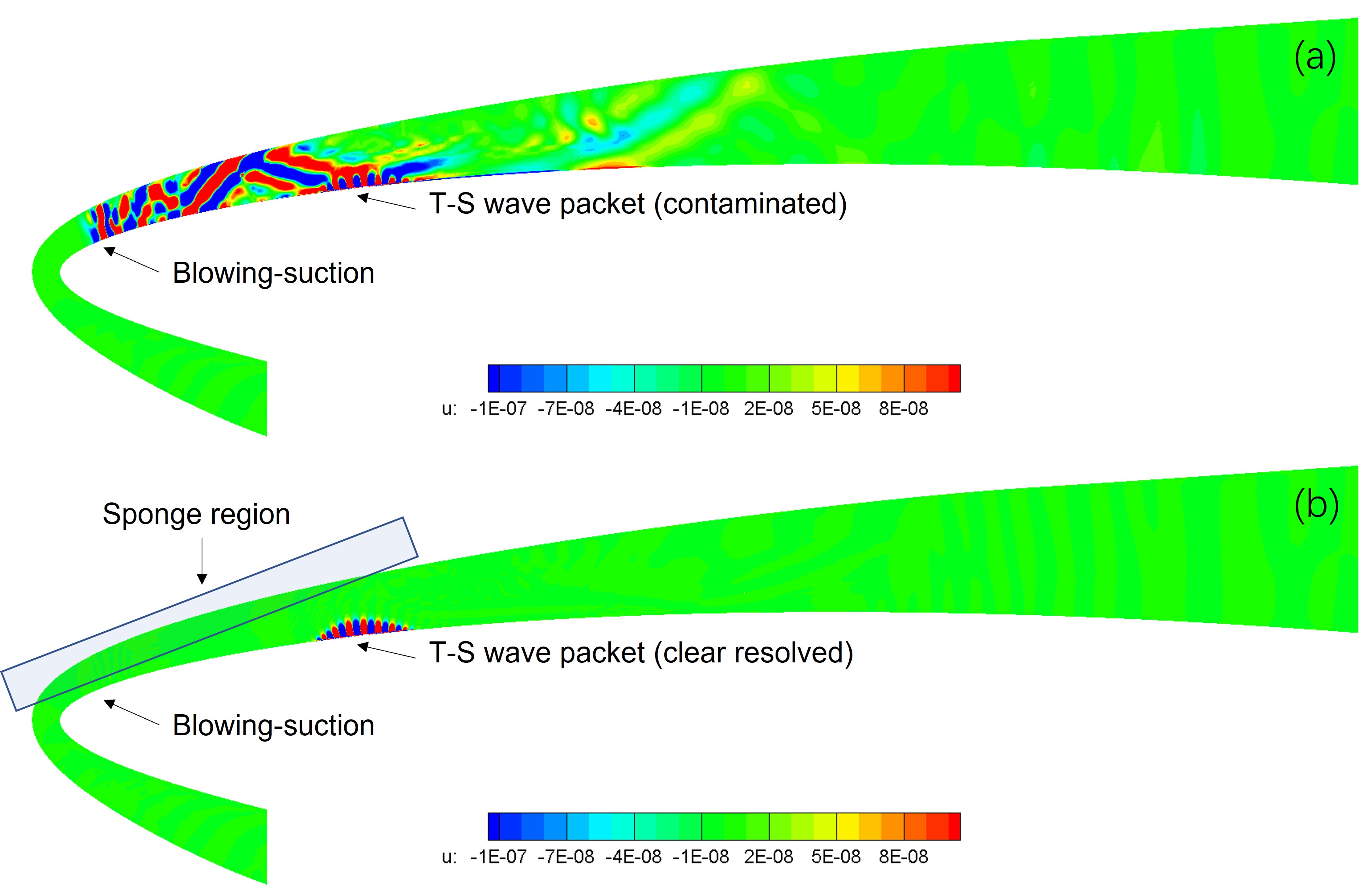}
  \caption{TS wave excitation over an airfoil in the reduced domain. (a) contaminated TS wave packet because of the wave reflection, and (b) clear resolved wave packet with the adoption of a sponge region.}
  \label{fig_workflow_waveReflection_comparison}
\end{figure}

To trip the boundary layer instabilities blowing and suction are introduced on the wall to at prescribed frequencies. This boundary perturbation not only causes the disturbance inside the boundary layer but also generate waves propagating that proaogate in all directions. If the inflow boundary is reflective (e.g. the entropy-pressure compatible inflow), the waves are reflected back to the wall. Fig.  \ref{fig_workflow_waveReflection_comparison}(a) shows this type of contamination in a reduced domain over a wing section of the the CRM-NLF model, which is studied in detail in section \ref{sec::CRM}. In the case in Fig. \ref{fig_workflow_waveReflection_comparison}, the blowing-suction is turned off after the first cycle to better trace the wave reflection between the wall and inflow boundary. It is apparent that the reflected waves are of similar magnitude of the target TS wave packet, and therefore the TS signals are difficult to be separated. A solution to this issue is to use a sponge region, which adds additional forcing terms to damp the difference components with respect to a mean  or steady field. As shown in Fig.  \ref{fig_workflow_waveReflection_comparison}(b), the adoption of sponge region effectively removes the wave reflection and the TS wave packet is clearly resolved. In addition to the sponge region, mesh coarsening \cite{edelmann2014influence} and lower order polynomials can be used for the spectral elements outside the boundary layer region to reduce the resolution and therefore smear the waves as they travel across this region, further damping the reflected waves.

\begin{figure}[hbt!]
  \centering
  \includegraphics[width=10cm]{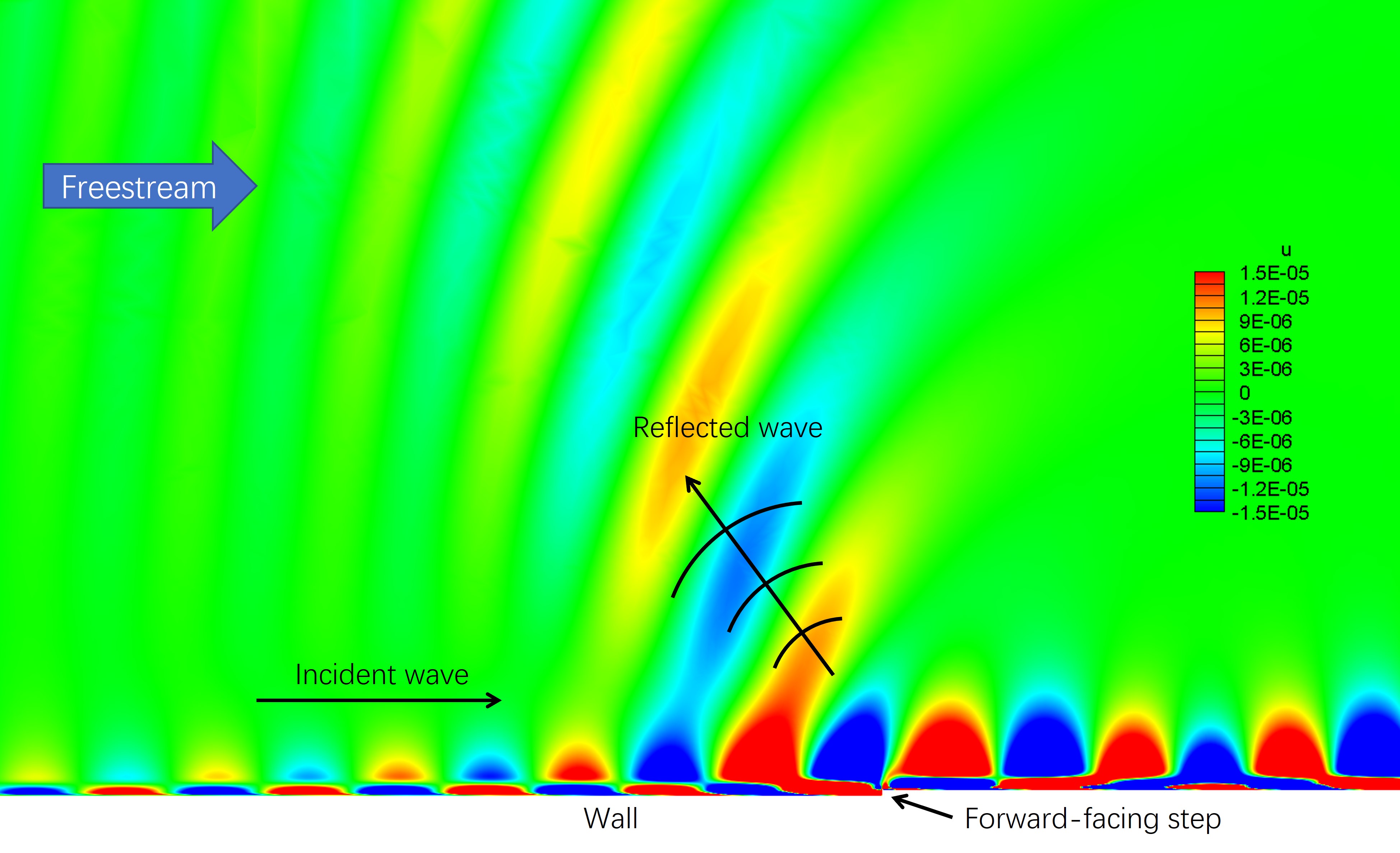}
  \caption{TS wave reflection at a forward-facing step.}
  \label{fig_workflow_waveReflection_FFS}
\end{figure}

As the excited waves travel downstream inside the boundary layer, if they experience a surface irregularity, waves are reflected as well. This leads to the second source of contamination. Fig. \ref{fig_workflow_waveReflection_FFS} provides an example for wave reflection at a forward-facing step, which will be introduced in section \ref{sec::flatPlate} as the verification of the workflow. In general these reflected waves are much weaker compared with the disturbances of interest in the most part of the domain. However, since the disturbances grow over several orders of magnitude as they travel downstream, and since the adoption of the $e^N$ method requires precise capture of the weakest amplitudes of the disturbances (which are used as the denominators in the calculation), the reflection waves can easily contaminate the results and lead to incorrect $N$-factor.

\begin{figure}[hbt!]
  \centering
  \includegraphics[width=0.49\textwidth]{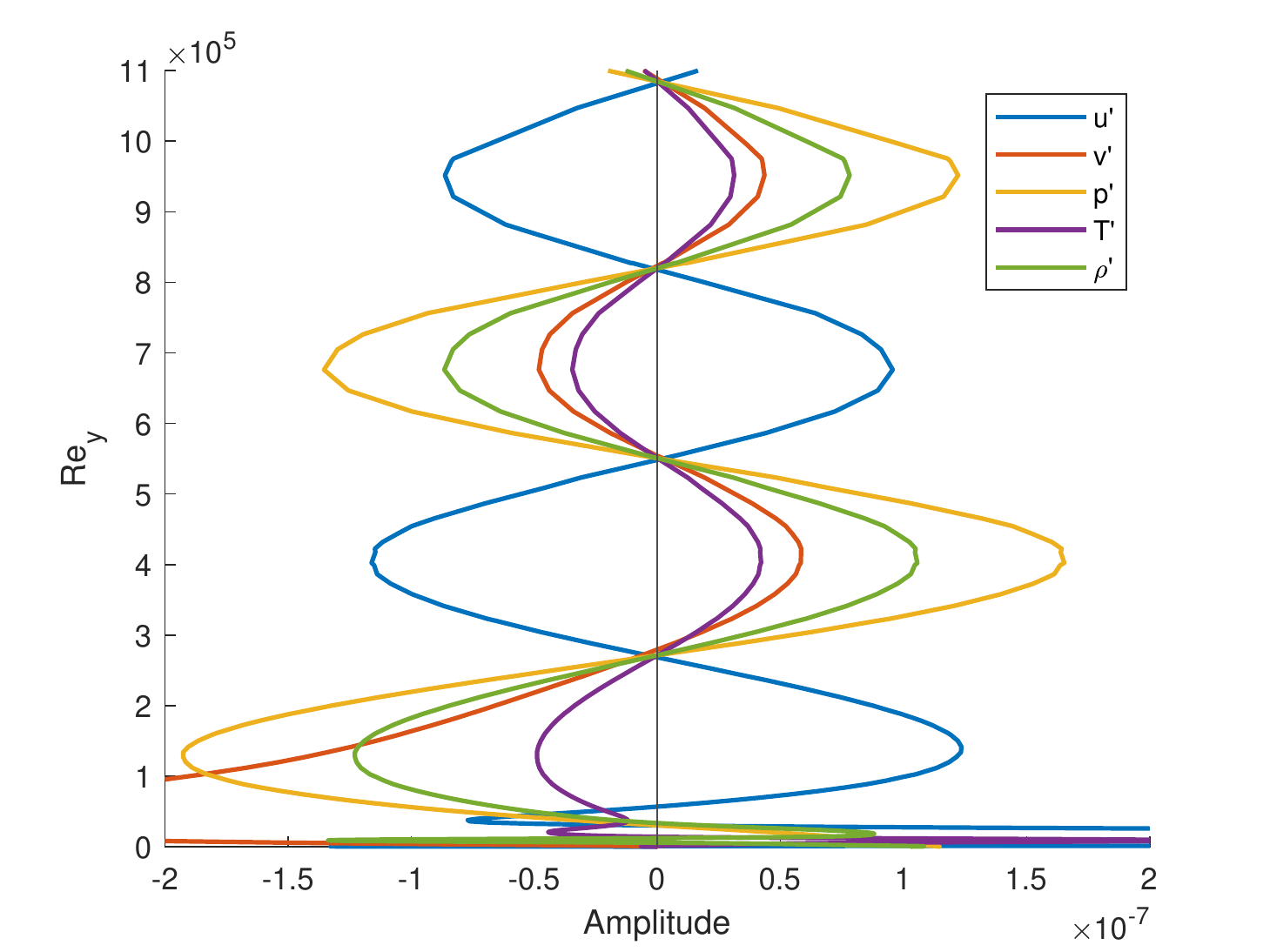}
  \caption{Reflected waves in the slice passing $Re_x=2.4177\times10^6$ on the wall and of $15$ deg with respect to the $y$-axis. (The forward-facing step is located at $Re_x=2.45\times10^6$.)}
  \label{fig_workflow_reflected_waves}
\end{figure}

To the best of the authors' knowledge, few methods are available to clean a contaminated signal, especially when the two signals have the same frequency (i.e. the incident wave and reflected waves are assume to have the same frequency). Schopper \cite{schopper1982analysis} spotted standing waves in the TS wave experiment and he explained the phenomena as the superposition of the TS waves and freestream acoustic waves. He modelled the acoustic field to obtain cleaner TS wave signals, where proper adjustment was still needed. However, it can be expected that acoustic modelling at the surface irregularities is more difficult than that for the freestream acoustics. Moreover, in the TS wave study by Edelmann \cite{edelmann2014influence} the reflected waves were considered as acoustics whereas the profile of the reflected waves of the same case (see section \ref{sec::flatPlate}) in  Fig. \ref{fig_workflow_reflected_waves} indicates a more complicated condition. The reflected waves are acoustics ($p'$) dominated but vorticity waves ($u'$ and $v'$) and entropy waves ($T'$ and $\rho'$) have comparable amplitude. ($u'$ has opposite phase to the others is also observed.) These complicated waves make the general wave modelling more difficult and therefore not an ideal method to be adopted.

\begin{figure}[hbt!]
  \centering
  \includegraphics[width=10cm]{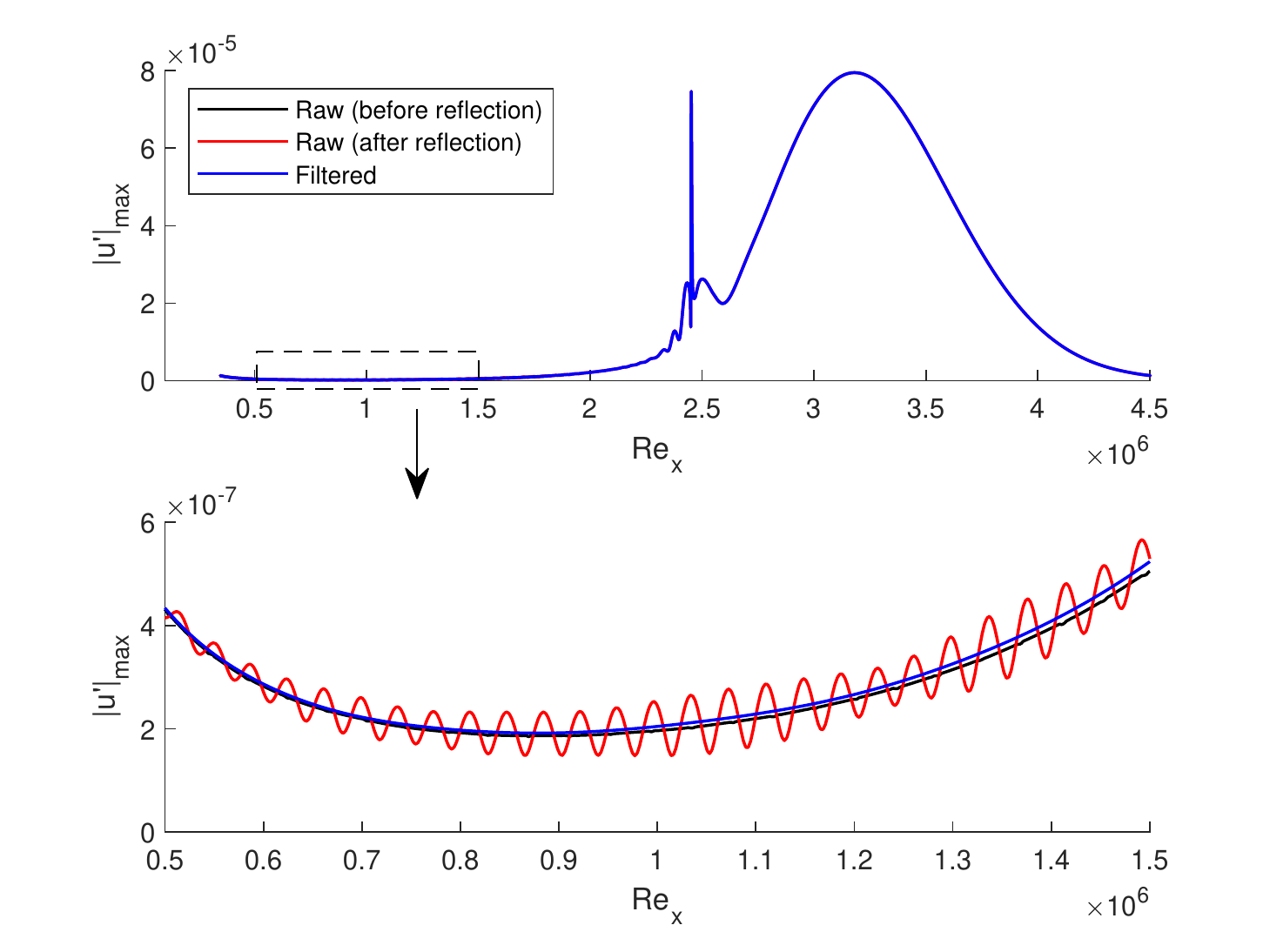}
  \caption{Comparison of raw data and the filtered results. The raw data is the amplification curve of 2D TS instability of $f=19$ kHz over the flat plate with a forward-facing step. $N=20$ is used to filter the data in the $Re_x$ interval $[3.45\times10^5,2.06\times10^6]$. Comparison of two sets of raw data and the filtered results. The first set of raw data (red line) includes oscillation because of the acoustic reflection at the forward-facing step. The second set of raw data (black line) is oscillation-free since it is obtained before the TS wavefront hits the step. The filtered data removes the oscillation using modified Chebyshev polynomials.}
  \label{fig_workflow_filtered_waves_comparison}
\end{figure}

To obtain the clean TS wave packet signal Edelmann used a “Moving Tukey Window” for Fourier transform to obtained the amplitude \cite{edelmann2014influence}. However, the moving speed needs to be known in advance to apply the window, which restricts the application of this method. To enable the amplitude of disturbances to be obtained in a user-friendly way, in the current workflow a signal filtered by modified Chebyshev polynomials \cite{boyd2001chebyshev} is adopted to remove the influence of the reflected waves. Fig. \ref{fig_workflow_filtered_waves_comparison} gives an example of the filtering, and compares the TS wave amplitude signal before reflection, with reflection, and the filtered result. The signal before reflection is taken before the TS wave reaches the step and therefore no reflected wave is generated. This signal can be considered as the ideal result although it is only available in a limited region upstream of the step. The signal with reflection is taken after both TS waves and the reflected waves have become fully developed in the domain. The filtered signal is post-processed based on the signal with reflection. The comparison shows that the relative difference between the minimum amplitude of the filtered signal and the ideal signal is only $2.68\%$, indicating the filtering effectively removes the contamination and recovers the signal to a desired level.


%%%%%%%%%%%%%%%%%%%%%%%%%%%%%%%%%%%%%%%%%%%%%%%%%%%%%%%%%%%%%%%%
\section{Linear growth limit and precision requirement for disturbance field computation}
\label{sec::precision}

In the $e^N$ method an empirical threshold for transition sits in the range of $6$--$11$. This is not as trivial as it sounds since it means a disturbance can grow up to $e^{11}$ ($\simeq 60000$) times or even higher whereas the amplitude of the disturbance still needs to be small enough so that the non-linearity can be neglected. Edelmann and Rist \cite{edelmann2015impact} obtained satisfactory results using DNS method by keeping the amplitude below $1 \times 10^{-4}$ of the freetream velocity in the whole domain, and a $N$-factor up to $9$ is presented. This corresponds to the smallest amplitude of $1.2 \times 10^{-8}$ of the freetream value. In the related work by Zahn and Rist \cite{zahn2016impact}, a maximum $N$-factor of $10.2$ is reported and the amplitudes of disturbances are roughly $1\times10^{-8}$ of the freestream value at introduction, which leads to a maximum non-dimensionalized amplitude of $2.7\times10^{-4}$.

However, the disturbance of order $O(10^{-4})$ may not be the limit to adopt the $e^N$ method together with DNS simulations. To figure out the capability for simulating the linear growth of disturbance in the current framework, the comparisons are made by introducing disturbance of different strength into a two dimensional flat plate boundary layer with a forward-facing step (see \ref{sec::flatPlate} for the geometry and detailed settings). The freestream Mach number is $0.8$ and the disturbance frequency of $19$ kHz is used to obtain representative results.

\begin{table}[hbt!]
  \centering
  \caption{Initial disturbance scaled by the freestream velocity.}
  \label{tab_cases_linearLimit_precision}
  \begin{tabular}{c c c c c c}
  \toprule
        & Case 1 &  Case 2 & Case 3 & Case 4 & Case 5\\  
  \midrule
  Initial disturbance & $5\times10^{-9}$ & $1\times10^{-7}$ & $1\times10^{-6}$ & $1\times10^{-5}$ & $1\times10^{-4}$\\
  \bottomrule
  \end{tabular}
\end{table}

\begin{figure}[hbt!]
    \centering
    \begin{subfigure}[b]{0.49\textwidth}
        \centering
        \includegraphics[width=\textwidth]{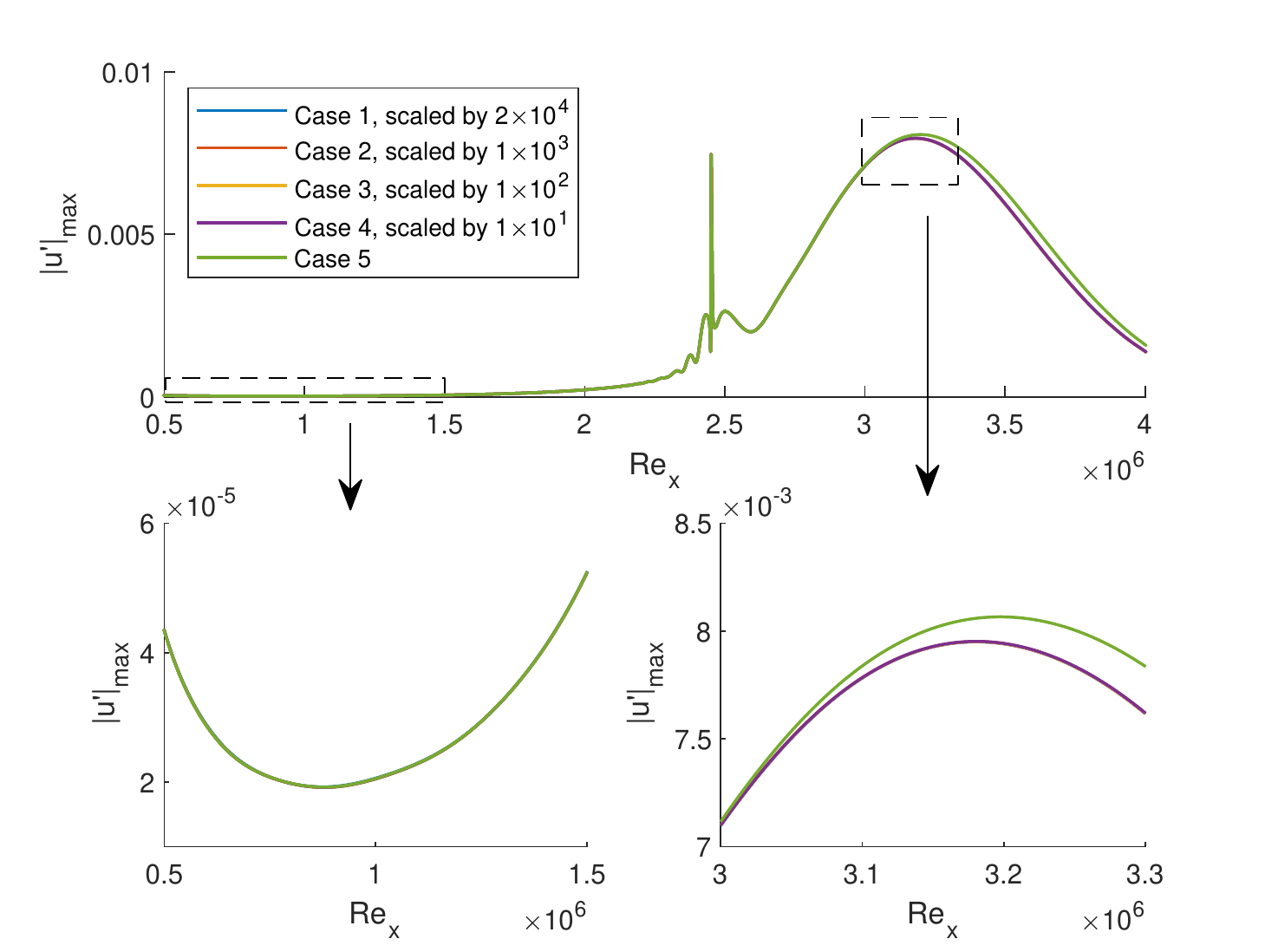}
        \caption{Maximum amplitude $|u'|_{max}$}
        \label{fig_linearLimit}
    \end{subfigure}
    \hfill
    \begin{subfigure}[b]{0.49\textwidth}
        \centering
        \includegraphics[width=\textwidth]{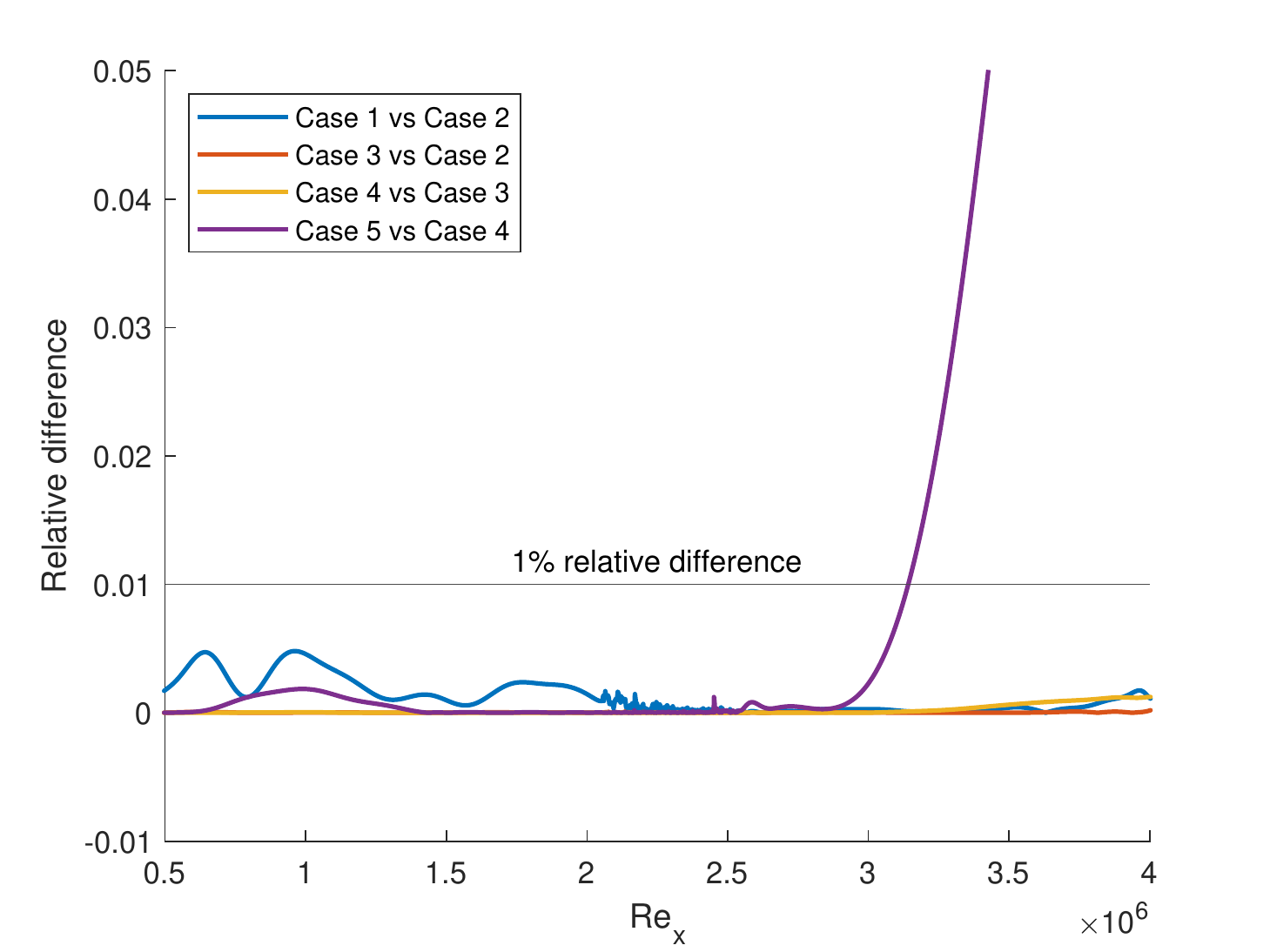}
        \caption{Relative difference of the scaled maximum amplitude}
        \label{fig_linearLimit_relativeDifference}
    \end{subfigure}
    \caption{Comparison of scaled TS wave growth curves of the five cases in Table \ref{tab_cases_linearLimit_precision}. The relative difference for each curve in (b) is computed using the adjacent two cases in Table \ref{tab_cases_linearLimit_precision}. The data in the latter case in the legend is used as denominator.}
    \label{fig_linearLimit_full}
\end{figure}

We compare five cases  as shown in Table \ref{tab_cases_linearLimit_precision}, where the  initial disturbances vary from weak to strong  with approximate magnitudes of  $5\times10^{-9}$, $1\times10^{-7}$, $1\times10^{-6}$, $1\times10^{-5}$, and $1\times10^{-4}$ of the freestream value. Fig. \ref{fig_linearLimit} provides the scaled curves for maximum amplitude of streamwise velocity of the excited TS wave in the streamwise direction. It shows that all the cases have a good linear scaling relation except Case 5, which has a larger amplification than the others near the peak. This larger amplification is likely arising from non-linear effects. The relative differences for the adjacent two cases are plotted in Fig. \ref{fig_linearLimit_relativeDifference}, where the relative difference for Case a and Case b, for example, is computed by
\begin{equation}
  \eta_{a,b}  = \frac{|(|u'|_{max}^a -c_{a,b} \cdot |u'|_{max}^b)|}{c_{a,b} \cdot |u'|_{max}^b}
\end{equation}
where $c_{a,b}$ is the scaling coefficient for the two cases, computed by the ratio of the maximum values of $|u'|_{max}^a$ and $|u'|_{max}^b$. It is shown that only the relative difference for Case 5 and Case 4 goes beyond $1$\% from $Re_x=3.14\times10^6$, corresponding to a $|u'|_{max}$ of $7.9\times10^{-3}$ of the freestream value in Fig. \ref{fig_linearLimit}. Therefore $7.9\times10^{-3}$ of the freestream value is considered as the estimated upper limit for linear growth.

\begin{figure}[hbt!]
    \centering
    \begin{subfigure}[b]{0.49\textwidth}
        \centering
        \includegraphics[width=\textwidth]{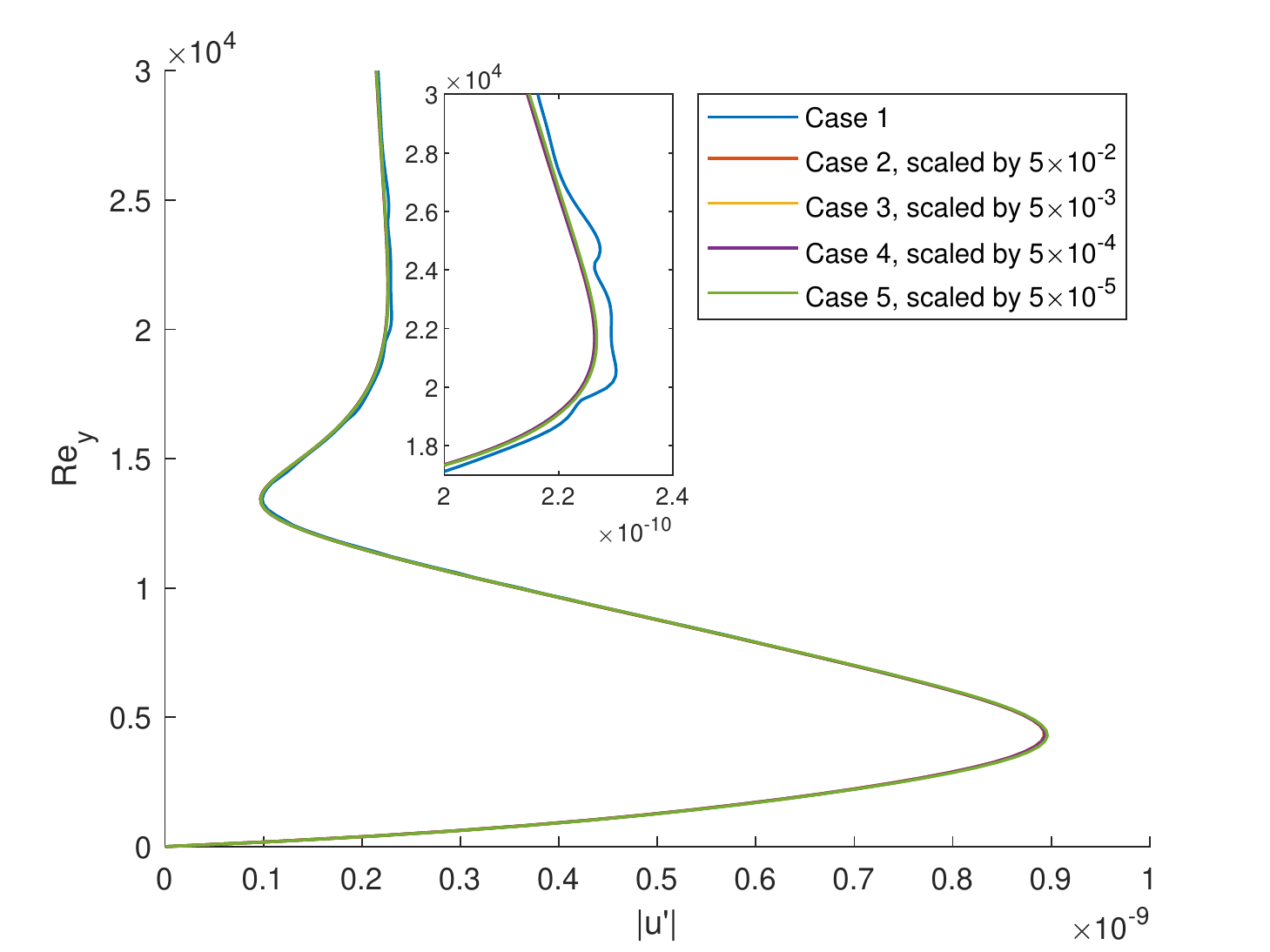}
        \caption{$|u'|$-profiles at $Re_x=8.35\times10^5$}
        \label{fig_precision_1Em9}
    \end{subfigure}
    \hfill
    \begin{subfigure}[b]{0.49\textwidth}
        \centering
        \includegraphics[width=\textwidth]{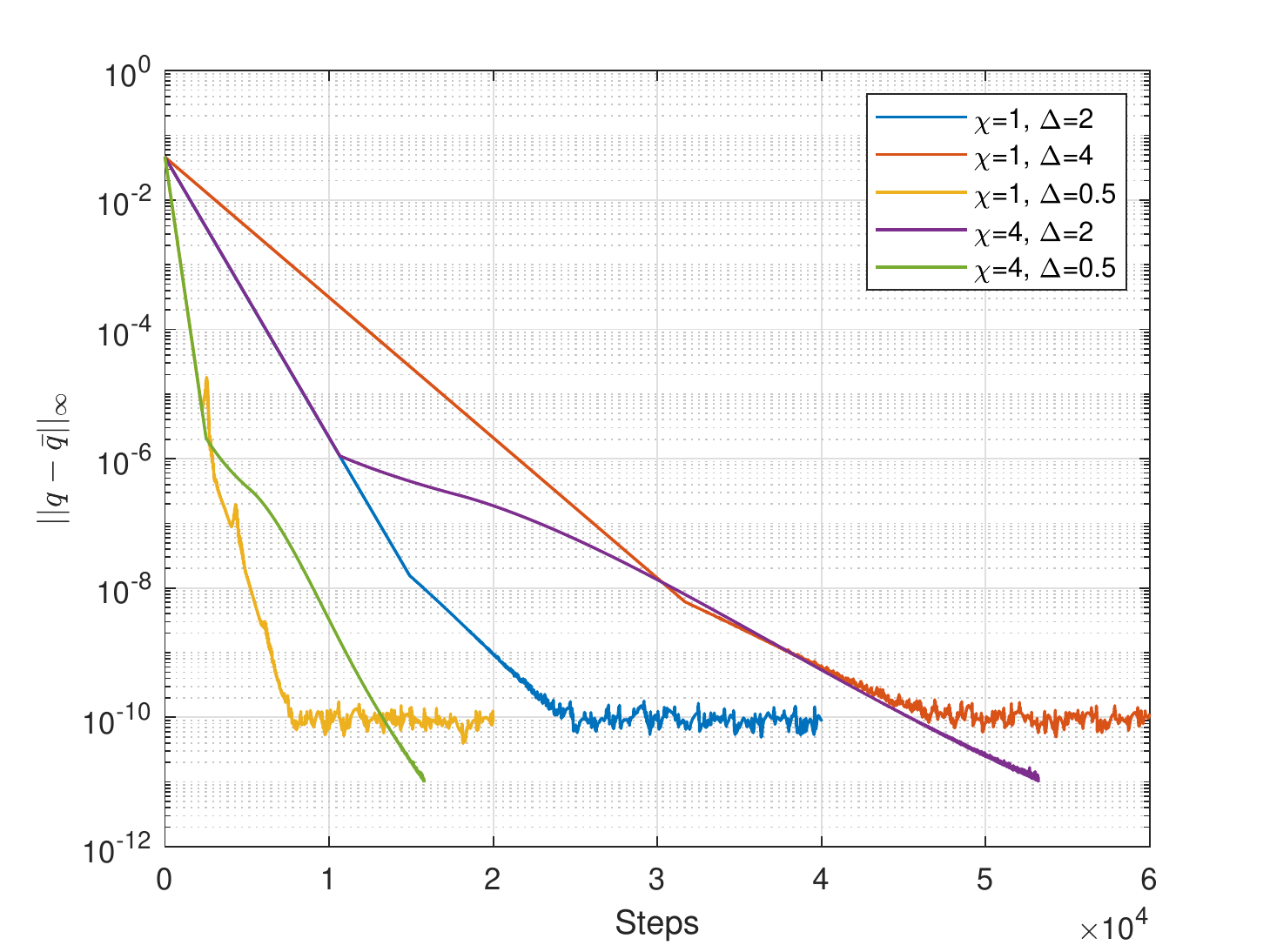}
        \caption{Baseflow convergence history using SFD.}
        \label{fig_FP_FFS_SFD_history}
    \end{subfigure}
    \caption{Precision verification: (a) the computational errors have the order of $O(10^{-12})$ -- $O(10^{-11})$; (b) the baseflow converges to $O(10^{-11})$.}
    \label{fig_precision}
\end{figure}

\begin{figure}[hbt!]
  \centering
  \includegraphics[width=8cm]{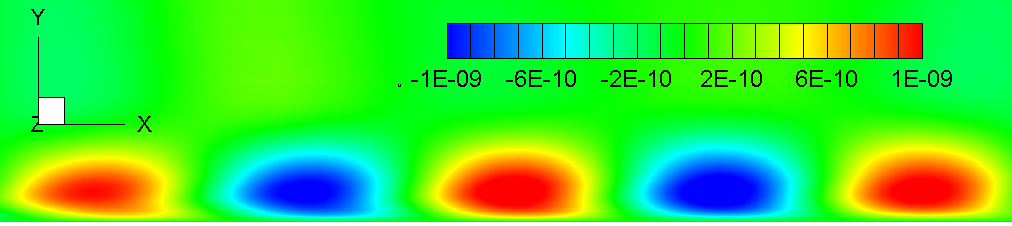}
  \caption{TS wave computation around $Re_x=8.35\times10^5$ in the presence of reflected waves. The $y$-axis is scaled by a factor of $5$.}
  \label{fig_FP_FFS_resolution_1Em9}
\end{figure}

On the other hand, an estimation for the lower limit is also needed since the disturbances whose amplitude smaller than the limit cannot not be well distinguished from the computational errors and undamped waves. The $|u'|$-profiles of these cases at $Re_x = 8.35\times10^5$ are plotted in Fig. \ref{fig_precision_1Em9}. This streamwise position is close to where the amplitudes of the disturbances reach their minimums. The figure shows that all of these profiles are well resolved and the computation errors cause a maximum relative error of $0.51$\% in the profile for Case 1. Since the amplitude of $1\times10^{-9}$ is well captured (from Case 1, also see Fig. \ref{fig_FP_FFS_resolution_1Em9}), this value can be used as the estimation for the lower limit. (Although the lower limit is not reached, for example, with respect to $1$\% relative error and smaller amplitude is still available, $1\times10^{-9}$ is enough for the estimation.) Therefore, the disturbances with amplitudes from $1\times10^{-9}$ to $7.9\times10^{-3}$ of the freestream value can be studied in a single simulation within the current framework. This gives the maximum $N$-factor of $15.9$, covering the typical transitional value from $6$ to $11$. For extreme cases where a larger $N$-factor is needed, the amplification curve for certain disturbance can be generated in two runs by placing the blowing-suction in two different streamwise position, and then combine the result in post-processing.

The above estimation is about the disturbances, which are obtained by subtracting the steady baseflow or mean flow from the perturbed fields. To make sure the disturbance of order of $O(10^{-9})$ is well resolved, the simulation of baseflow needs to converge to the order of $O(10^{-10})$ -- $O(10^{-11})$. This can be achieved by direct time marching for a long time or using the Selective Frequency Damping (SFD) for higher efficiency. In each step of the SFD, the current field is low-passed filtered (with width $\Delta$) to generate a predicted steady field, and the difference of the two fields is damped with a control coefficient $\chi$ \cite{jordi2014encapsulated}. The time stepping stops when the difference is lower than a user-specified threshold. For example, if a time-dependent system reads
\begin{equation}
  \label{eq::sfd_1}
  \dot{q} = g(q)
\end{equation}
the system with SFD takes the form
\begin{equation}
\label{eq::sfd_2}
\left\{
\begin{aligned}
  \dot{q}       & = g(q) - \chi (q-\bar{q}) \\
  \dot{\bar{q}} & = \frac{q - \bar{q}}{\Delta}
\end{aligned}
\right.
\end{equation}

A convergence verification for the baseflow for cases in Table \ref{tab_cases_linearLimit_precision} is given in Fig. \ref{fig_FP_FFS_SFD_history}. With the control coefficient $\chi=1$ the baseflow only converges to $1\times10^{-10}$ since some TS wave-shaped structures continue to exist downstream of the step whereas $1\times10^{-11}$ convergence is achieved by increasing the control coefficient to $4$.

%%%%%%%%%%%%%%%%%%%%%%%%%%%%%%%%%%%%%
\section{Workflow verification: TS wave development in 2D transonic flat plate boundary layer}
\label{sec::flatPlate}

In this section 2D TS wave development in two transonic flat plate boundary layer flows are simulated, and the $N$-factors are computed to the verify to the workflow. The two boundary layers are over a clean geometry and a geometry with a forward-facing step (FFS), respectively. These two cases have been studied by Edelmann and Rist \cite{edelmann2015impact} using the DNS as well, and their results are used for comparisons.

\begin{figure}[hbt!]
  \centering
  \includegraphics[width=10cm]{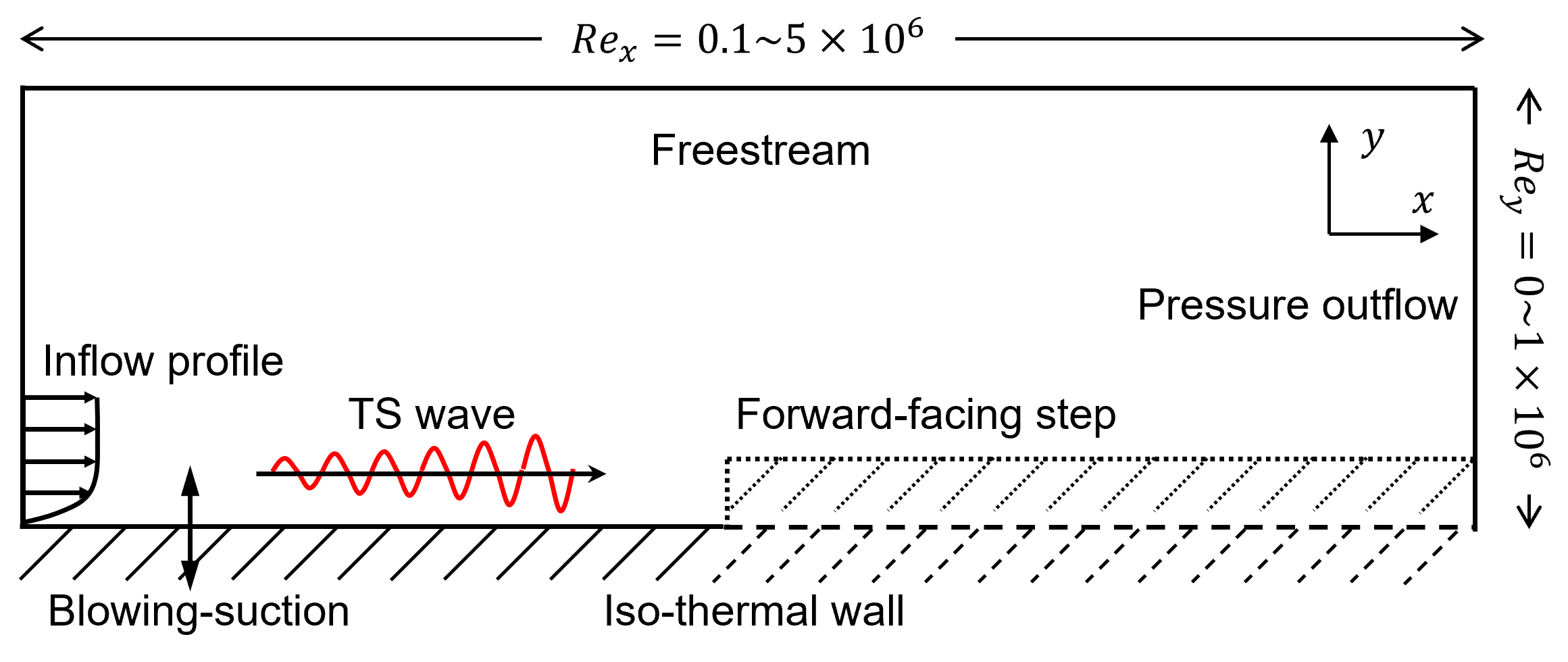}
  \caption{Schematic representation of the flow problem. The dashed line represents the wall for the clean geometry, and the dotted line represents the wall for the stepped geometry.}
  \label{fig_FP_geometry}
\end{figure}

Figure \ref{fig_FP_geometry} shows the geometries and computational domains for the two cases. For the stepped case, detailed parameters of the step are given in Table \ref{tab::FP2D_setting}, where $Re_S$ is the Reynolds number based on the step position, $Re_H$ is the Reynolds number based on the step height, $H/\Theta$ is the step height over momentum thickness at the very position in the clean case, $Re_{H,H}$ is the Reynolds number based on step height and velocity at the step height in the clean case, $L_1/H$ and $L_2/H$ are the separation region length upstream and downstream of the step over the step height, respectively. In our simulations the flow conditions for the two cases are exactly the same. Both of them have a freestream Mach number  $Ma=0.8$, freestream temperature $T_\infty=288$ K, and  Prandtl number $Pr=0.71$. The wall is iso-thermal with $T_{wall}=T_\infty$. A compressible boundary layer profile are weakly imposed at the inflow ($Re_x=1\times10^5$) while a pressure outflow with freestream value is adopted.
\begin{table}[hbt!]
  \centering
  \caption{Parameters for the forward-facing step on the plate.}
  \label{tab::FP2D_setting}
  \begin{tabular}{c c c c c c}
    \toprule
    $Re_S$ & $Re_H$ & $H/\Theta$ & $Re_{H,H}$ &$ L_1/H$ & $L_2/H$ \\  
    \midrule
    $2.45\times10^6$ & $2640$ & $\sim 2.4$ & $\sim 1350$ & $\sim 4.8$ & $\sim 0.9$\\
    \bottomrule
    \end{tabular}
\end{table}

\begin{figure}[hbt!]
    \centering
    \begin{subfigure}[b]{0.49\textwidth}
        \centering
        \includegraphics[width=\textwidth]{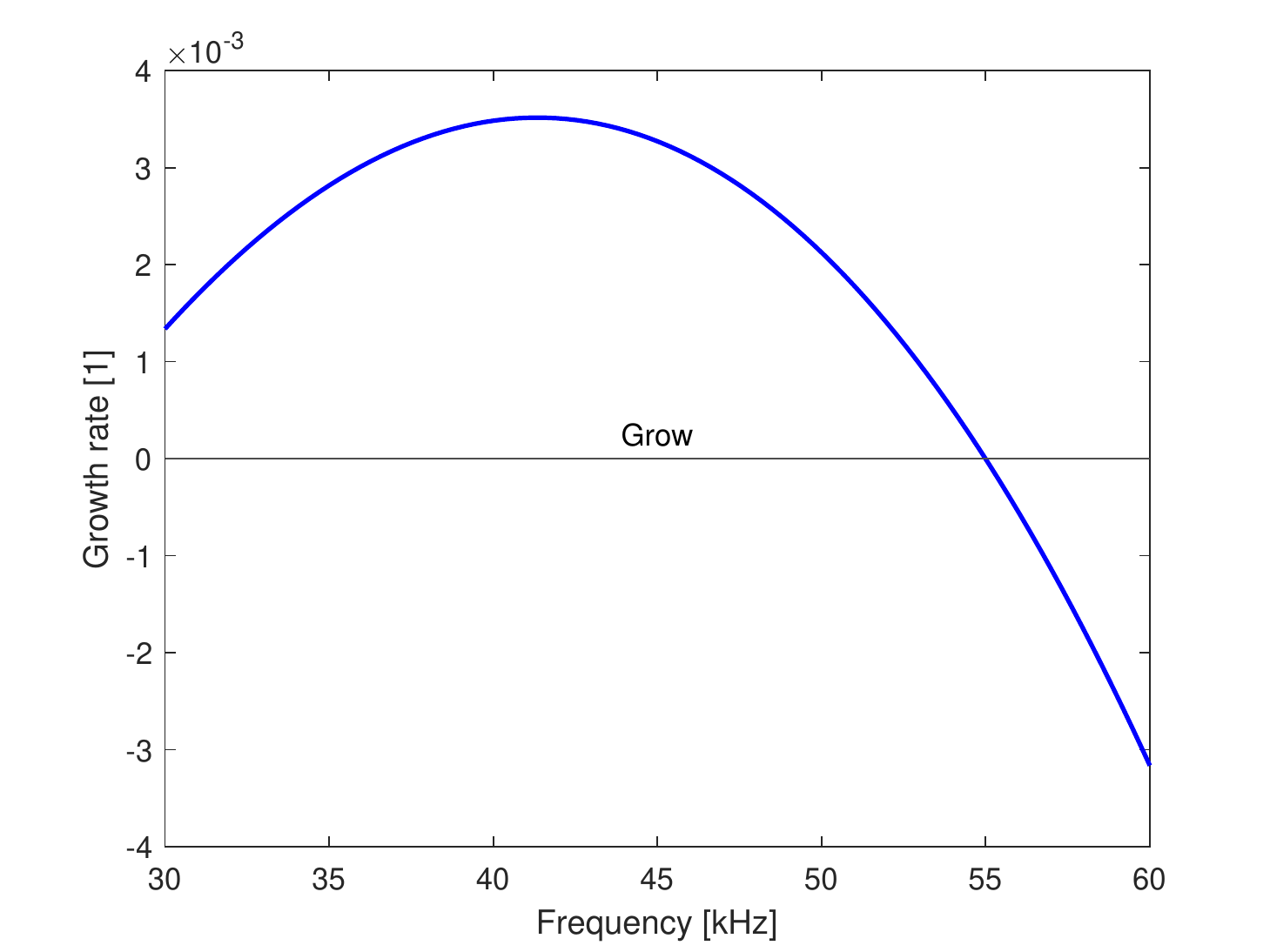}
        \caption{$x/L=10$\% ($Re_x = 0.59\times10^6$)}
        \label{fig_FP_freqScan:a}
    \end{subfigure}
    \hfill
    \begin{subfigure}[b]{0.49\textwidth}
        \centering
        \includegraphics[width=\textwidth]{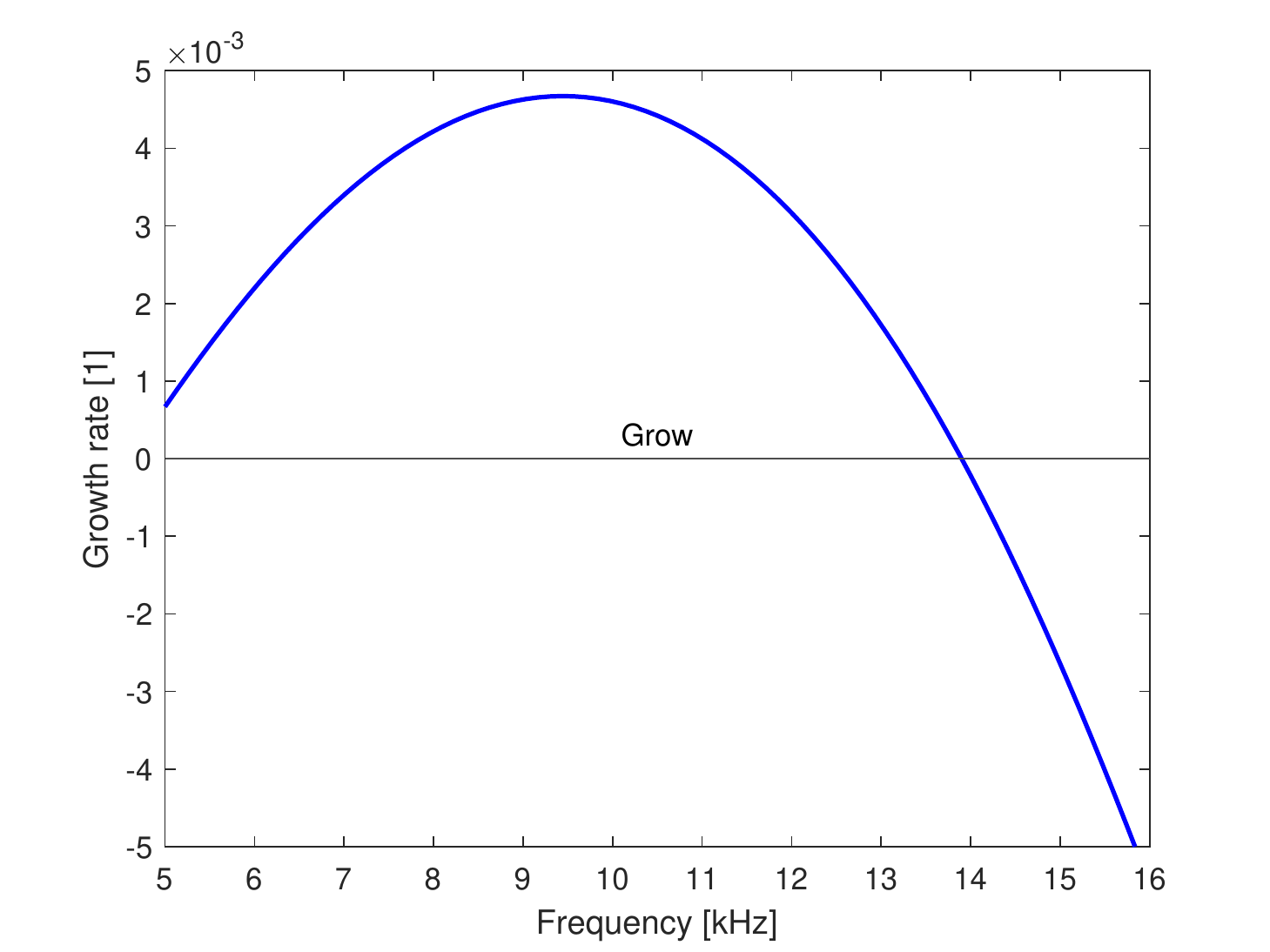}
        \caption{$x/L=90$\% ($Re_x = 4.51\times10^6$)}
        \label{fig_FP_freqScan:b}
    \end{subfigure}
    \caption{Frequency scans at an upstream position and a downstream position.}
    \label{fig_FP_freqScan}
\end{figure}

Since the geometry is simple, the background RANS simulation is skipped. We direct compute steady baseflows using the compressible flow solver in Nektar++. To find an appropriate frequency range for the TS waves to be simulated, sectional instability analysis by LST is performed using the boundary layer profile of the clean case. The profiles at $10$\% and $90$\% streamwise positions are analyzed. Frequency scans, where the grow rate of TS waves with a range of frequencies are computed, are performed and the results are provided Fig. \ref{fig_FP_freqScan}. It shows that the most amplified TS waves have the frequencies of approximately $42$ kHz and $10$ kHz at $x/L=10$\% and $x/L=90$\%, respectively. Therefore, with these reference frequencies, TS waves of $16$ different frequencies in the range $[13,45]$ kHz are selected for simulations. The same frequency range is used for both clean and stepped cases.

\begin{figure}[hbt!]
    \centering
    \begin{subfigure}[b]{0.49\textwidth}
        \centering
        \includegraphics[width=\textwidth]{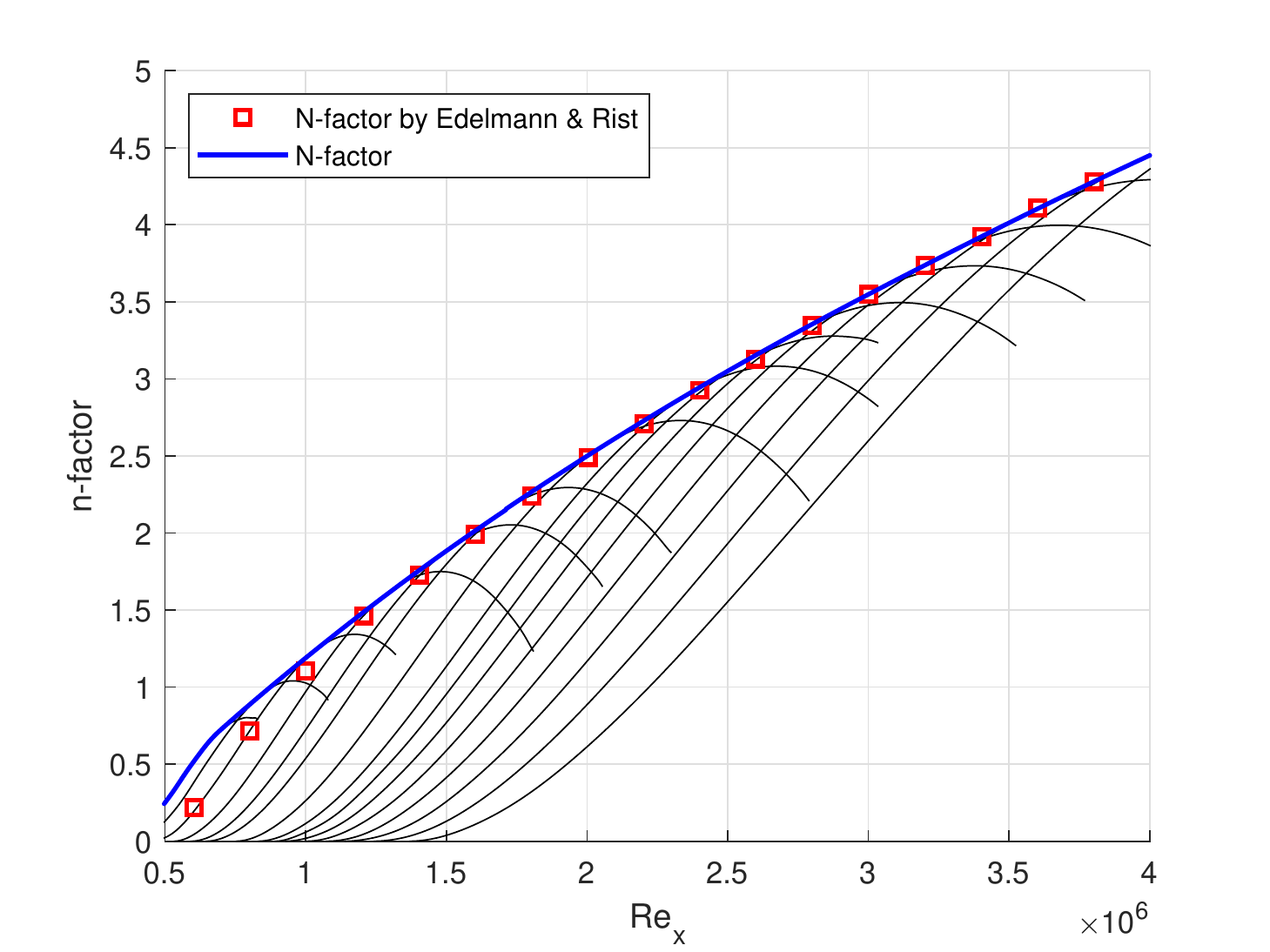}
        \caption{Clean flat plate}
        \label{fig_FP_clean_N}
    \end{subfigure}
    \hfill
    \begin{subfigure}[b]{0.49\textwidth}
        \centering
        \includegraphics[width=\textwidth]{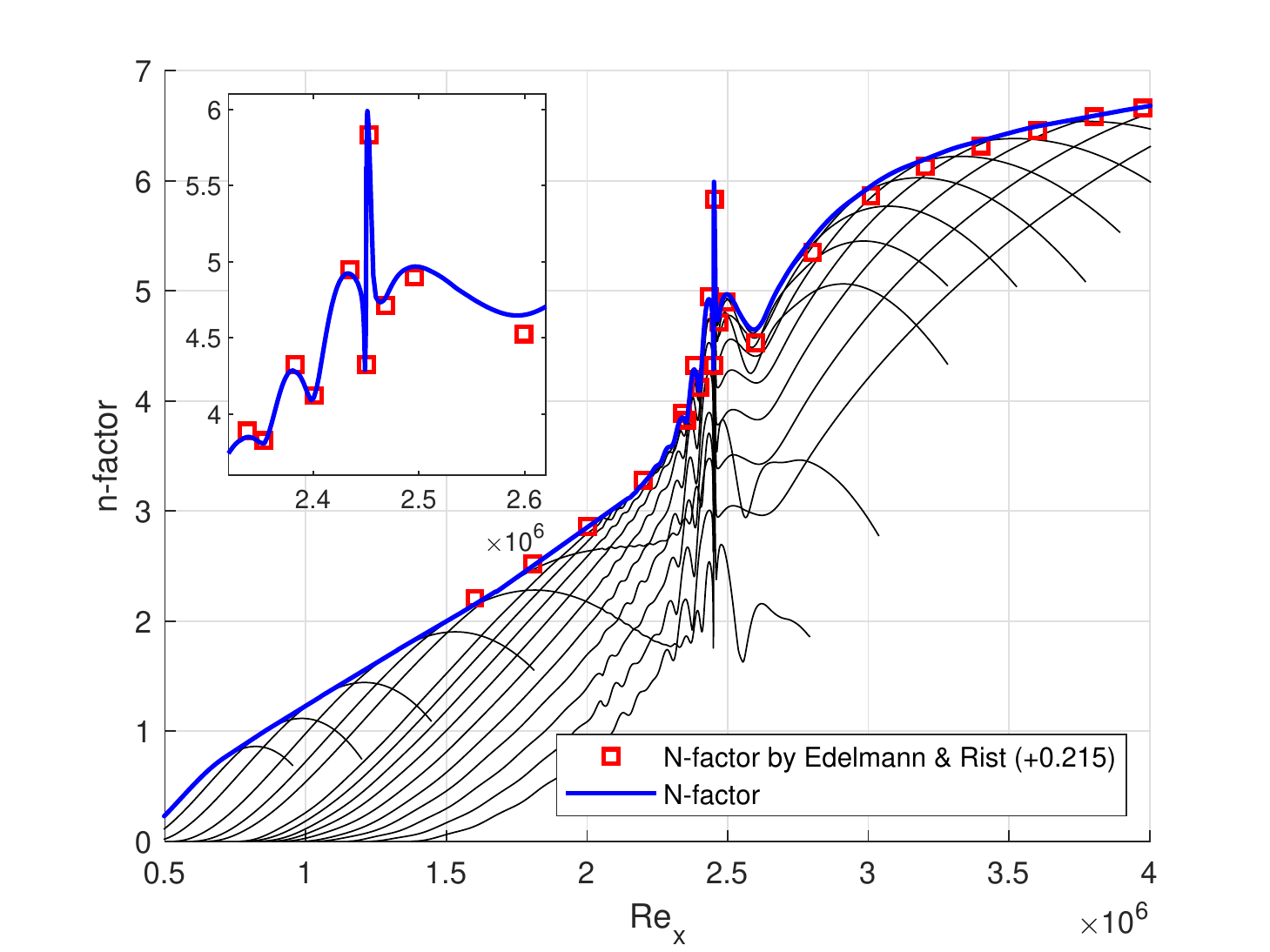}
        \caption{Flat plate with a forward-facing step}
        \label{fig_FP_FFS_N}
    \end{subfigure}
    \caption{$N$-factors for the two flat plate cases.}
    \label{fig_FP_N}
\end{figure}

\begin{figure}[hbt!]
  \centering
  \includegraphics[width=0.49\textwidth]{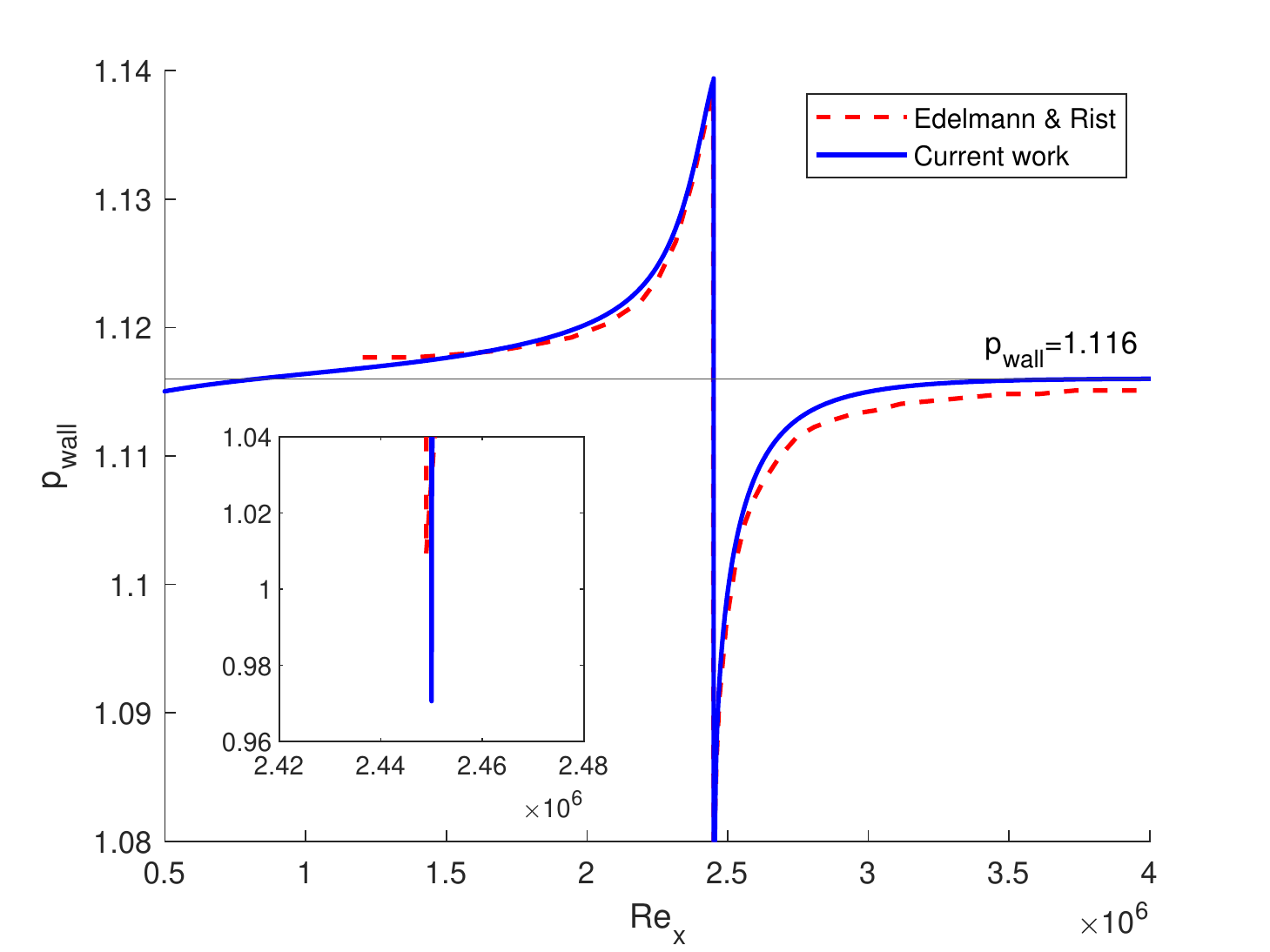}
  \caption{Wall pressure distribution comparison.}
  \label{fig_FP_FFS_pwall}
\end{figure}

Figure \ref{fig_FP_clean_N} and \ref{fig_FP_FFS_N} show the final $N$-factor curves for the two case as well as the component $n$-factors. It is apparent that the $N$-factor for clean case well agrees with the data from the reference for $Re_x>1.2\times10^6$. The discrepancy at low $Re_x$ is due to the different domains size and treatments to high-frequency TS waves. The higher $N$-factor in the current work comes from further upstream domain being included in the simulations whereas it is assumed to be truncated in the reference work. As for the comparison for the stepped case, good agreement is also achieved after shifting the reference data by $+0.215$ in the plot. The peaks and troughs of the oscillations near the FFS also agrees well with the reference. The above agreements indicate the development of TS waves are  simulated sufficiently well, and the data are correctly obtained from the simulations.  The shift in the reference data is needed due to the differences in the baseflow. As shown in Fig. \ref{fig_FP_FFS_pwall}, the wall pressure distributions in the current work and the reference work are not perfectly matched. This is because in the presence of the step, the baseflow is unavoidably modulated and the zero pressure gradient condition does not hold when using the same boundary conditions as the clean case, particularly when the inflow profile is weakly enforced. Compared with the reference, a larger adverse pressure gradient upstream of the step is observed while the pressure gradient downstream is slightly smaller. It is these different pressure gradients that leads to a larger growth in the $N$-factor in Fig. \ref{fig_FP_FFS_N} since the development of TS wave is sensitive to pressure distribution and is destabilized by the adverse pressure gradient.


%%%%%%%%%%%%%%%%%%%%%%%%%%%%%%%%%%%%%%%%%%%%%%%%%%%%%%%%%%%%%%%%%%%%
\section{Transitional study of transonic boundary layer flow over CRM-NLF model}
\label{sec::CRM}

With the workflow verified, in this section we analyze the transitional performance of a wing section of the open-accessed NASA Common Research Model with Natural Laminar Flow wing (CRM-NLF) \cite{crm_nlf}. The CRM-NLF wing is designed to have a flat upper surface to avoid strong pressure gradient and significant amplification of crossflow waves. We therefore focus on the growth of TS waves.

\begin{table}[hbt!]
  \centering
  \caption{Conditions for the RANS simulation of the CRM-NLF model.}
  \label{tab_freestream_conditions}
    \begin{tabular}{c c c c c c c c}
    \toprule
    $Ma$ & $Re_L$ & $\alpha$ [deg] & $T_{\infty}$ [K] & $U_{\infty}$ [m/s] & $\rho_{\infty}$ [kg/m$^3$] & $L$ [m]\\  
    \midrule
    $0.856$ & $8.5\times10^6$ & $1.5$ & $277.1$ & $285.7$ & 3.343 & 0.154678\\
    \bottomrule
    \end{tabular}
\end{table}

\begin{figure}[hbt!]
    \centering
    \begin{subfigure}[b]{0.45\textwidth}
        \centering
        \includegraphics[width=\textwidth]{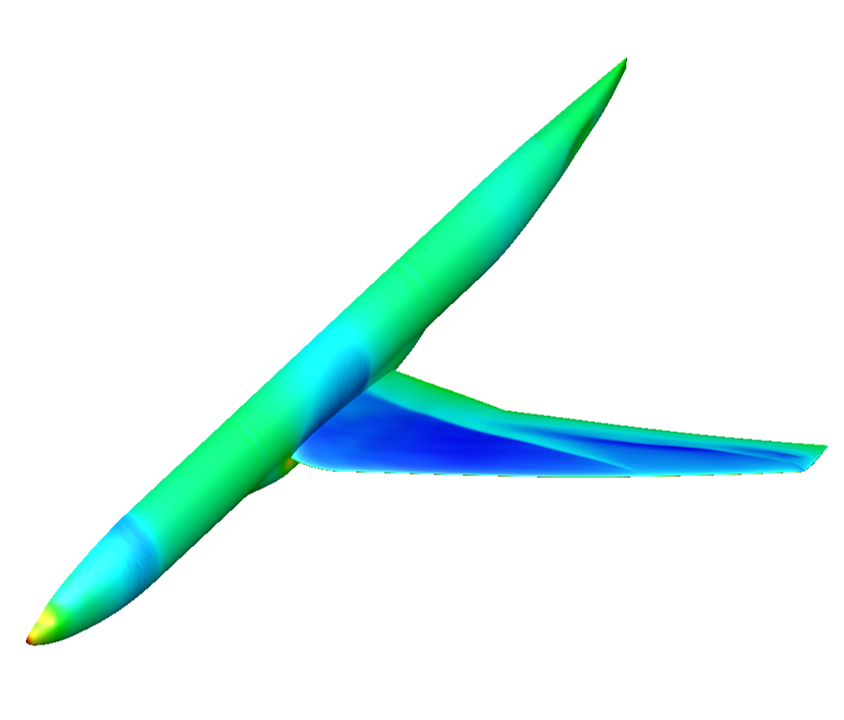}
        \caption{Pressure distribution by the RANS simulation.}
        \label{fig_CRM_RANS}
    \end{subfigure}
    \hfill
    \begin{subfigure}[b]{0.45\textwidth}
        \centering
        \includegraphics[width=\textwidth]{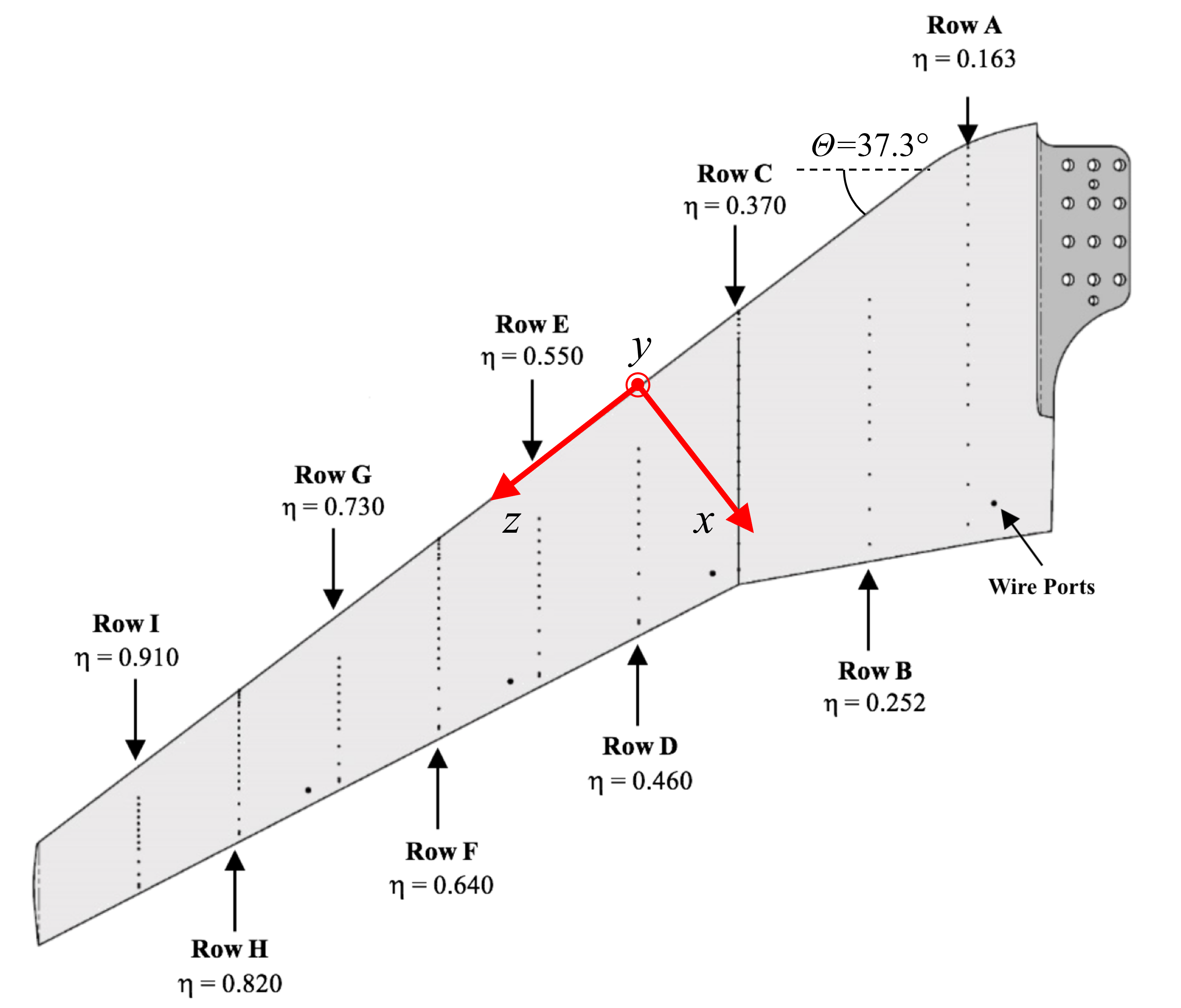}
        \caption{Wing section and the coordinate system}
        \label{fig_CRM_section}
    \end{subfigure}
    \caption{Background RANS simulation and coordinate system for simulation in the reduced domain.}
    \label{fig_CRM_RANS_section}
\end{figure}

A background field is first generated through a RANS simulation over the full geometry at the transonic freestream conditions in Table. \ref{tab_freestream_conditions}. The pressure distribution on the wing surface is shown in Fig. \ref{fig_CRM_RANS}, where two shocks on the wing can be spotted. Accordingly, the reduced domain is set up so that the outflow boundary is located upstream of the shock, since downstream of the shock the flow is unlikely to keep laminar due to shock-induced transition and further analysis is less meaningful. As shown in Fig. \ref{fig_CRM_section}, the reduced domain is set on the slice normal to the leading-edge of the wing, and passing the leading-edge of Row D. The $x$-axis is in the normal-to-leading edge direction, the $y$-axis is in the vertical direction, and $z$-axis is in the spanwise direction.

\begin{figure}[hbt!]
  \centering
  \includegraphics[width=10cm]{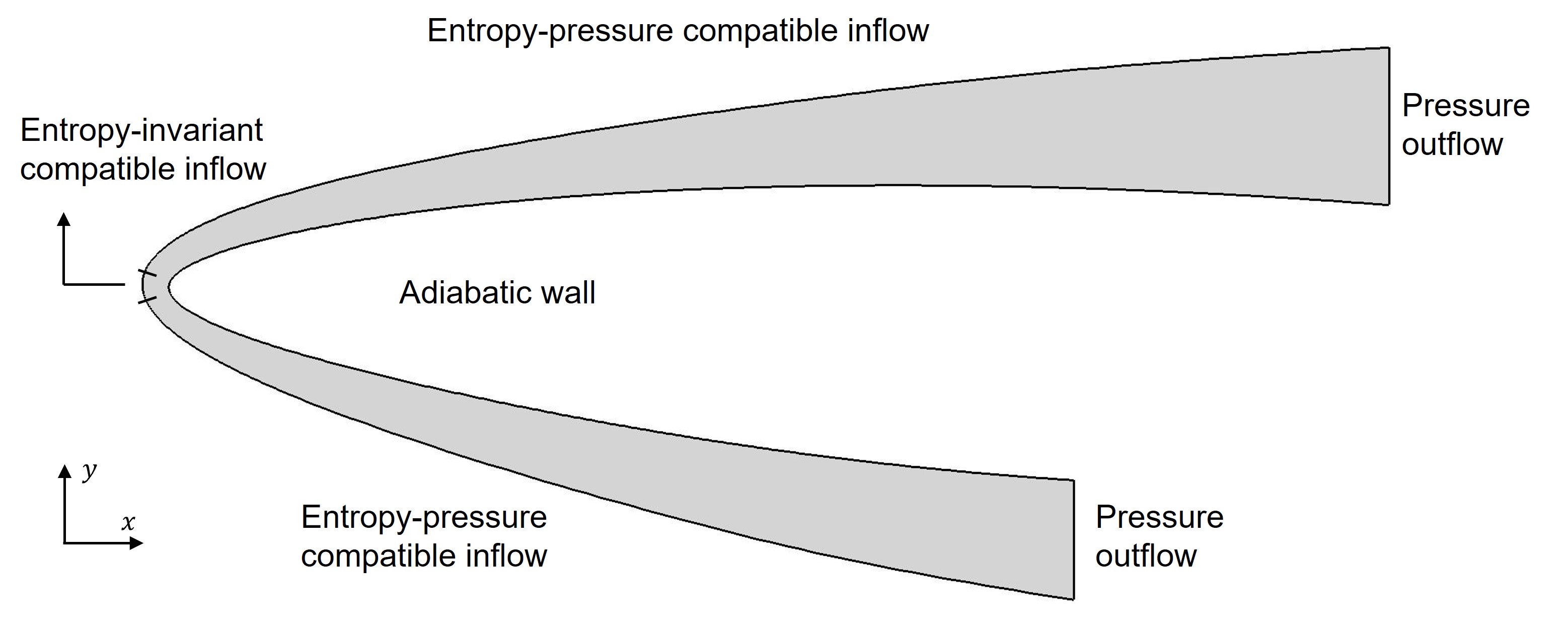}
  \caption{Boundary conditions for CRM-NLF case. Periodic condition is adopted in the $z$-direction for a quasi-3D simulation.}
  \label{Fig_airfoil_BCs}
\end{figure}

Fig. \ref{Fig_airfoil_BCs} gives the geometry of the reduced domain together with the boundary condition strategy for advection terms. As derived in section \ref{sec::pressure_compatible_inflow}, the entropy-pressure compatibility is mainly enforced in the inflow boundary, however  the entropy-invariant compatible inflow (the standard Riemann inflow) is applied at the nose region of the domain, covering the streamlines that pass the stagnation point, to achieve a well-pose problem. In addition, to carrying out a 3D simulation, periodic condition are adopted in the $z$-direction while for a 2D simulation the $w$-component velocity as well as the corresponding dynamic energy are removed from the interpolated RANS data. All other boundary conditions are imposed as explained in Ref. \cite{mengaldo2014guide}.

\begin{figure}[hbt!]
    \centering
    \begin{subfigure}[b]{0.38\textwidth}
        \centering
        \includegraphics[width=\textwidth]{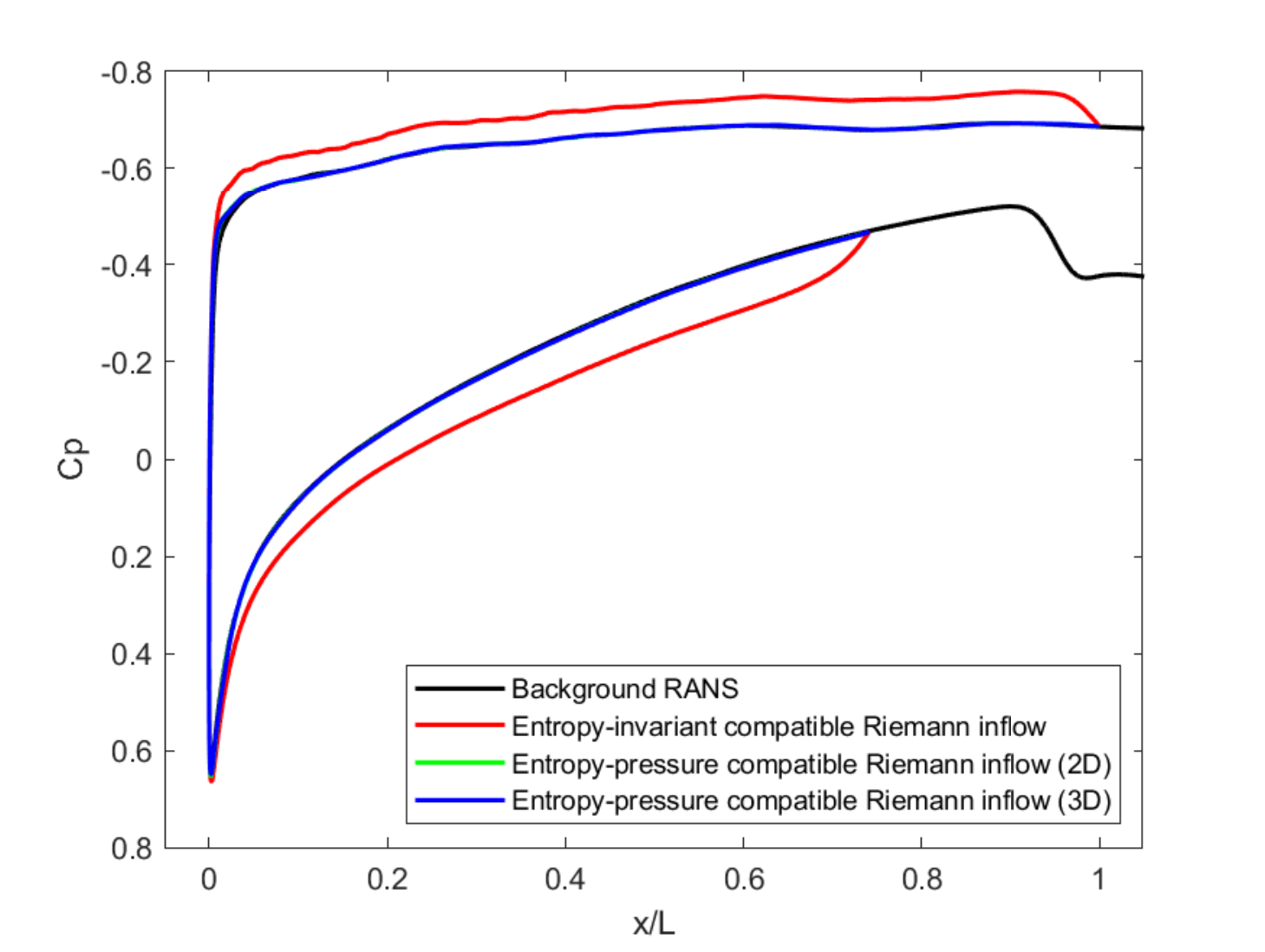}
        \caption{$C_p$ distribution}
        \label{fig_CRM_Cp}
    \end{subfigure}
    \hfill
    \begin{subfigure}[b]{0.3\textwidth}
        \centering
        \includegraphics[width=\textwidth]{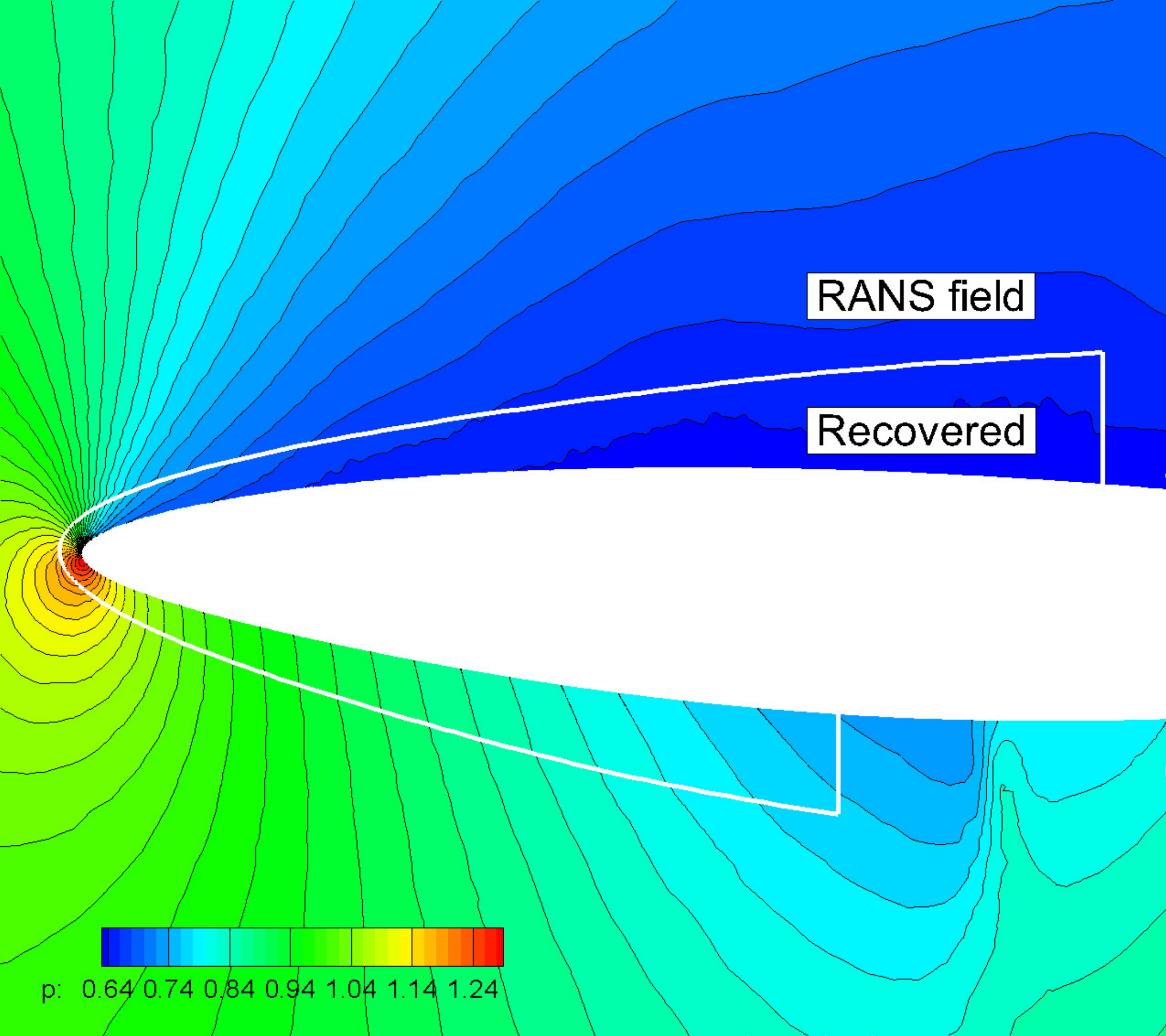}
        \caption{Pressure contour plot}
        \label{fig_CRM_p}
    \end{subfigure}
    \hfill
    \begin{subfigure}[b]{0.3\textwidth}
        \centering
        \includegraphics[width=\textwidth]{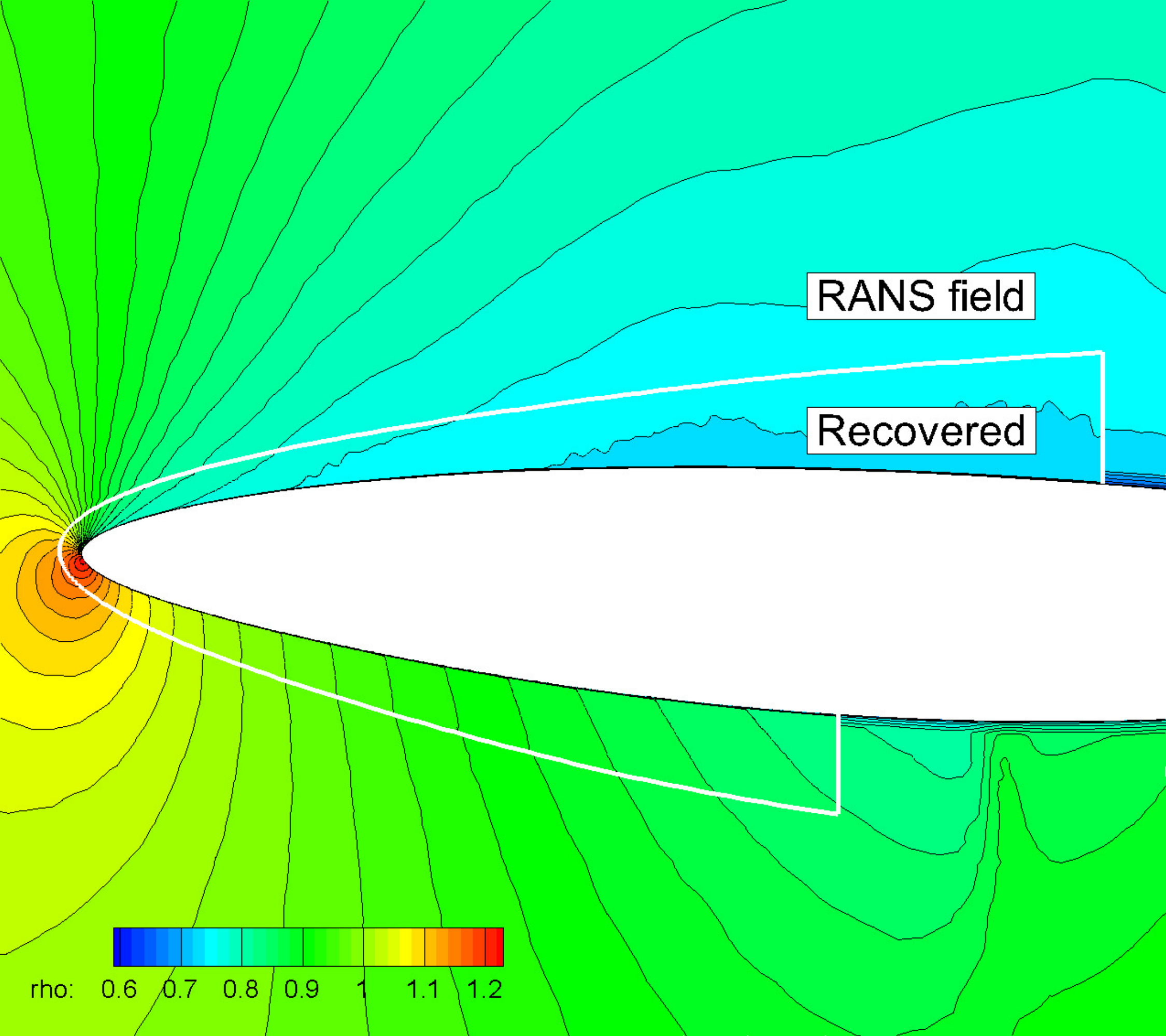}
        \caption{Density contour plot}
        \label{fig_CRM_rho}
    \end{subfigure}
    \caption{Baseflow results comparison with the outer RANS data.}
    \label{fig_CRM_baseflow}
\end{figure}

Before considering the amplification results, we first demonstrate the performance of the entropy-pressure compatible inflow by comparing it with the entropy-invariant compatible inflow. The baseflow pressure coefficient ($C_p$) distributions with adoption of these two inflow conditions are plotted in Fig.\ref{fig_CRM_Cp}. In the result for entropy-invariant inflow, significant deviations from the RANS data are observed on both sides of the wing section. These deviations are potentially caused by the use the reduced domain. The background RANS simulation is carried out the full complex geometry, where the wing-fuselage interaction, three-dimensionality, and the shock can influence the distributions. However, in the reduced domain only the wing section geometry and distributions along the boundaries are taken into consideration. The lack of complex geometry together with the transonic characteristic feature requires the boundary conditions to be carefully designed to recover the pressure fields of the background RANS simulation. As for the $C_p$ distributions for the both 2D and 3D baseflows using the entropy-pressure inflow, they  show excellent agreement with the RANS data, indicating the effectiveness of this boundary condition enforcement. Fig. \ref{fig_CRM_p} and \ref{fig_CRM_rho} further compare the pressure and density fields the in the reduced domain and the outer RANS fields, where we observe that the contour lines are well matched.

\begin{figure}[hbt!]
    \centering
    \begin{subfigure}[b]{0.33\textwidth}
        \centering
        \includegraphics[width=\textwidth]{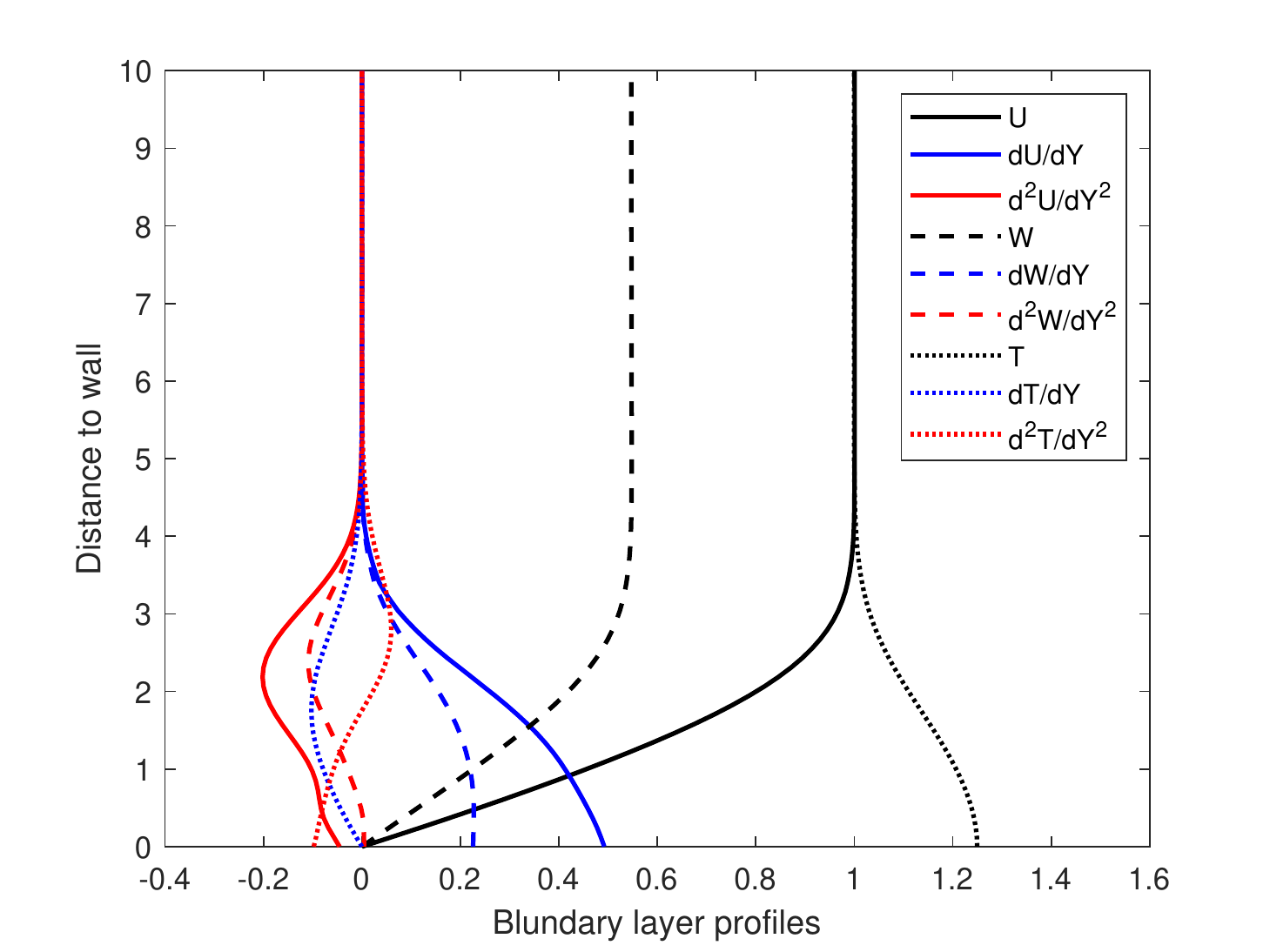}
        \caption{Boundary layer profiles}
        \label{fig_profile}
    \end{subfigure}
    \hfill
    \begin{subfigure}[b]{0.33\textwidth}
        \centering
        \includegraphics[width=\textwidth]{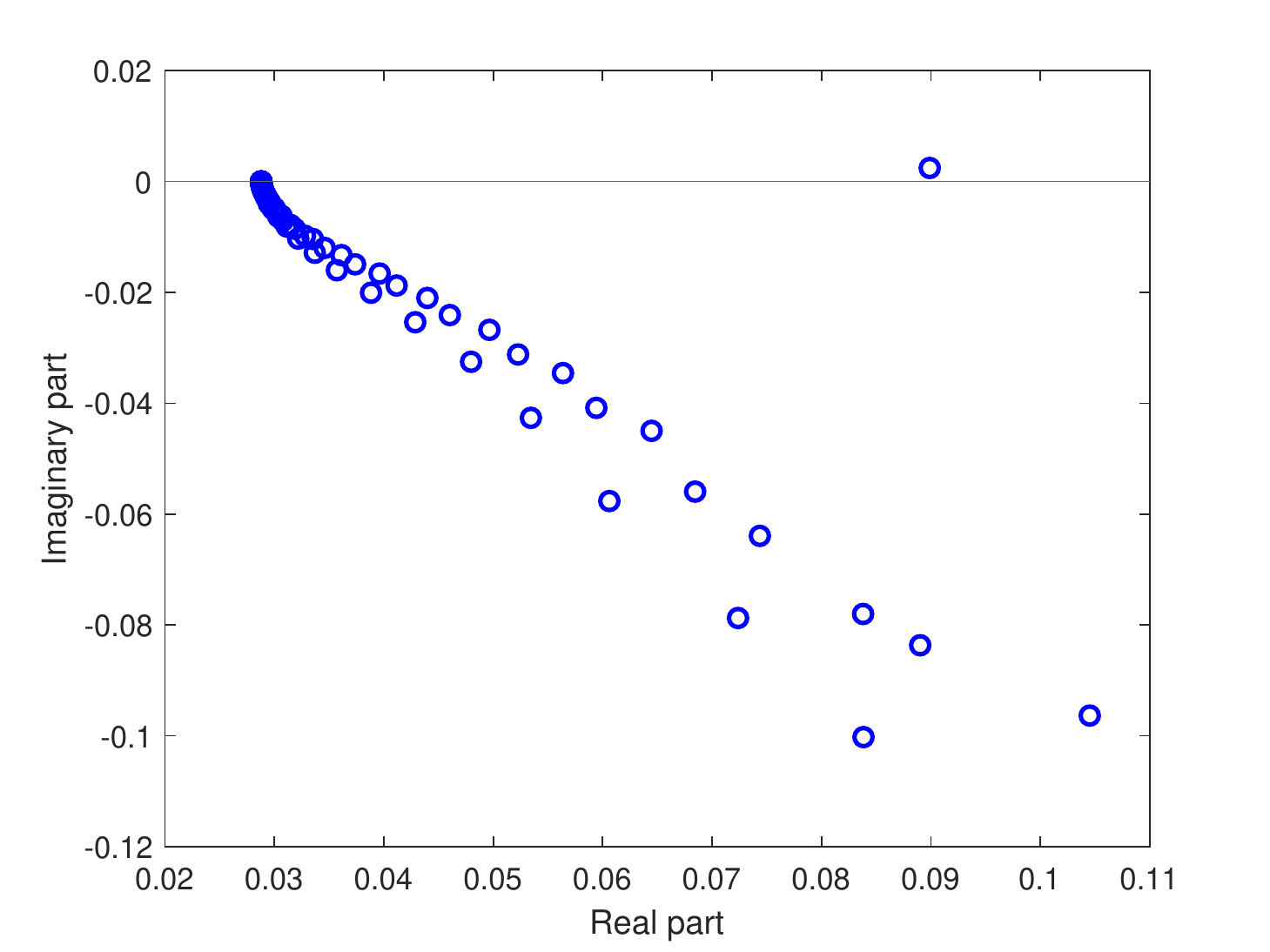}
        \caption{Eigenspectrum }
        \label{fig_spectra}
    \end{subfigure}
    \hfill
    \begin{subfigure}[b]{0.33\textwidth}
        \centering
        \includegraphics[width=\textwidth]{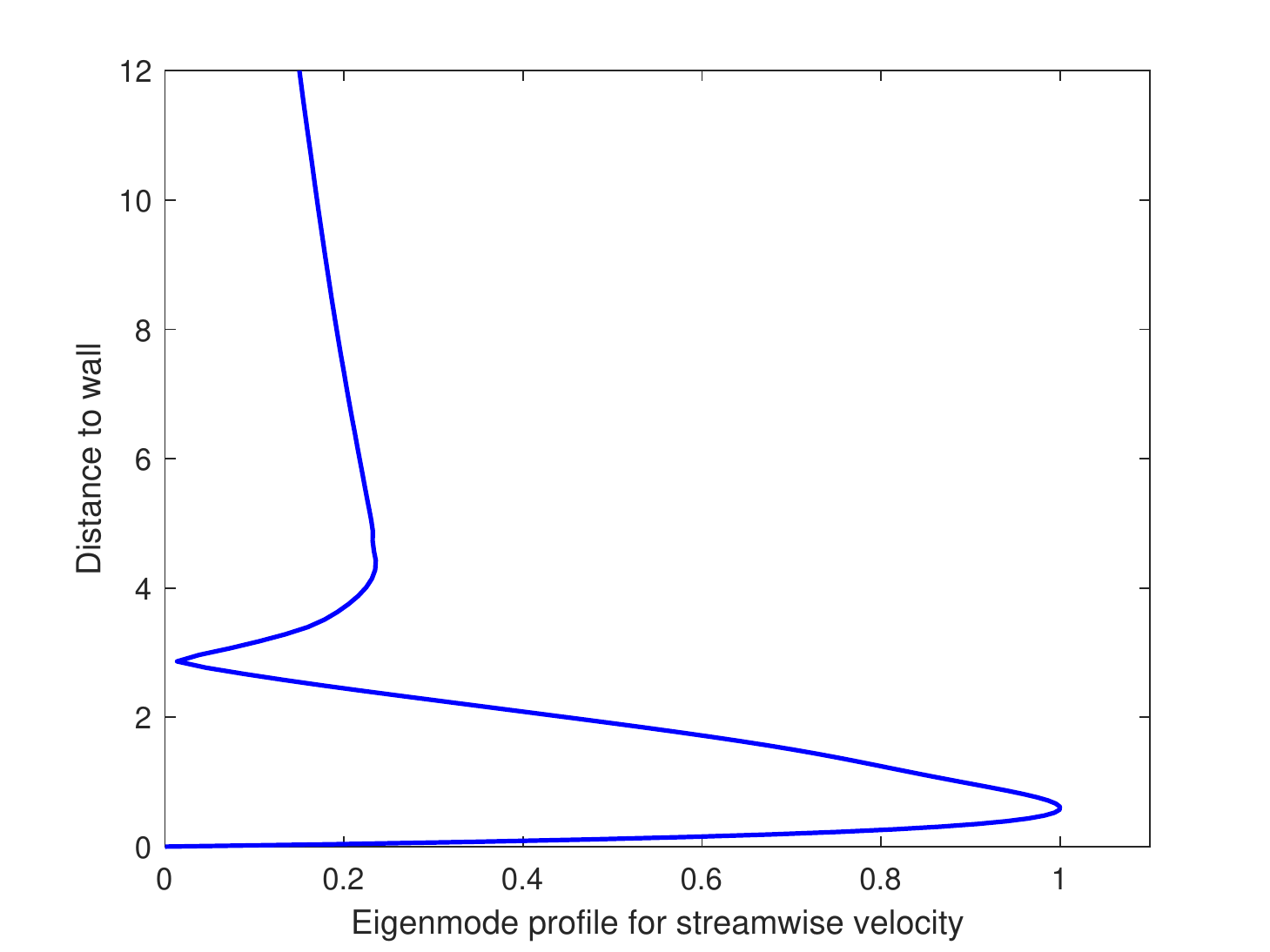}
        \caption{Eigenmode}
        \label{fig_eigenmode}
    \end{subfigure}
    \caption{Sectional stability analysis at $x/L=50\%$. Frequency $f = 30$ kHz and spanwise wavenumber $\beta = 100$ 1/m are used.}
    \label{fig_CRM_lst}
\end{figure}

After considering the baseflow computation we not turn our attention to the sectional stability analysis, which is performed in several streamwise positions. As the first step, the boundary layer profiles for velocity and temperature are extracted in the wall-normal direction, and their first and second order derivatives with respect to the wall-normal coordinate are computed using finite difference. When computing the second order derivatives, the non-smoothness originated from the data and the interpolation-induced noises could destabilize the following eigenvalue analysis or introduce pseudo-modes. The second order derivatives are therefore smoothed using modified Chebyshev polynomials. The scaled boundary layer profiles at $x/L=50\%$ is provided in Fig. \ref{fig_profile} as an example. The normal velocity and its derivatives are not shown since they are not needed according to the LST (and also small). The sectional stability analysis is performed based on the profile data. It first computes the eigenspectrum, where the pseudo-mode together with acoustic modes are automatically removed for a clean spectrum, as shown in Fig. \ref{fig_spectra}. In the spectrum the most possible mode is selected while the wave angle and eigenfunction should be double checked to make sure it is a real TS mode as given in Fig. \ref{fig_eigenmode}. This TS mode is then used as the initial condition for a spanwise wavenumber-frequency scan, where the growth rate of the wave are computed with respect to a specified a range of spanwise wavenumbre and frequency in a marching way. Therefore, if the initial condition fails to be the target mode, e.g. a mode belonging to the continues branch is mis-selected, the scan would generate incorrect results. The scan results inform the user which disturbances are the most energized at a given section. It is noted that all of the above operations can be finished within the NekPy interface. 

\begin{figure}[hbt!]
    \centering
    \begin{subfigure}[b]{0.33\textwidth}
        \centering
        \includegraphics[width=\textwidth]{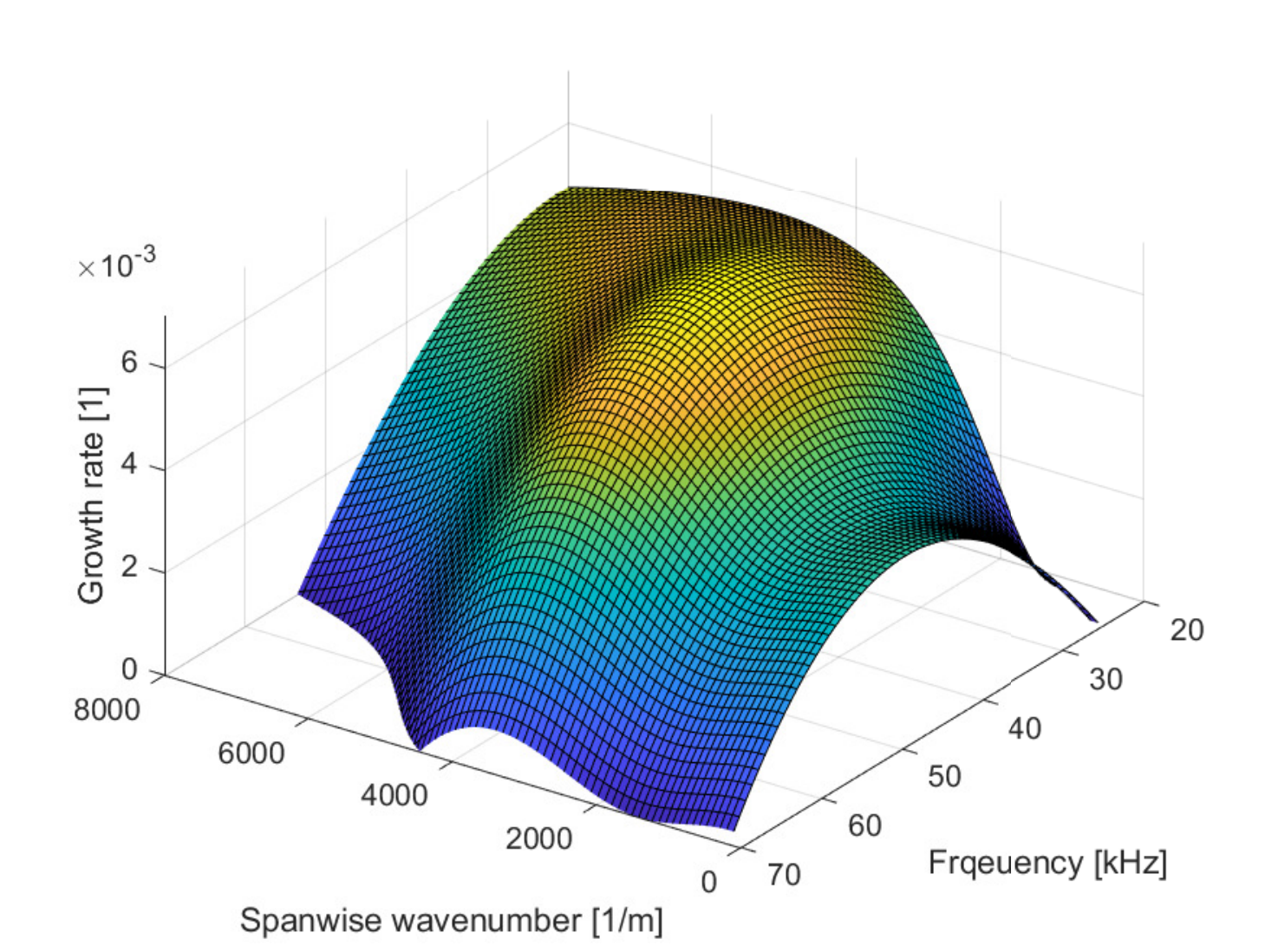}
        \caption{$x/L=30\%$}
        \label{fig_CRM_3D_scan_x03}
    \end{subfigure}
    \hfill
    \begin{subfigure}[b]{0.33\textwidth}
        \centering
        \includegraphics[width=\textwidth]{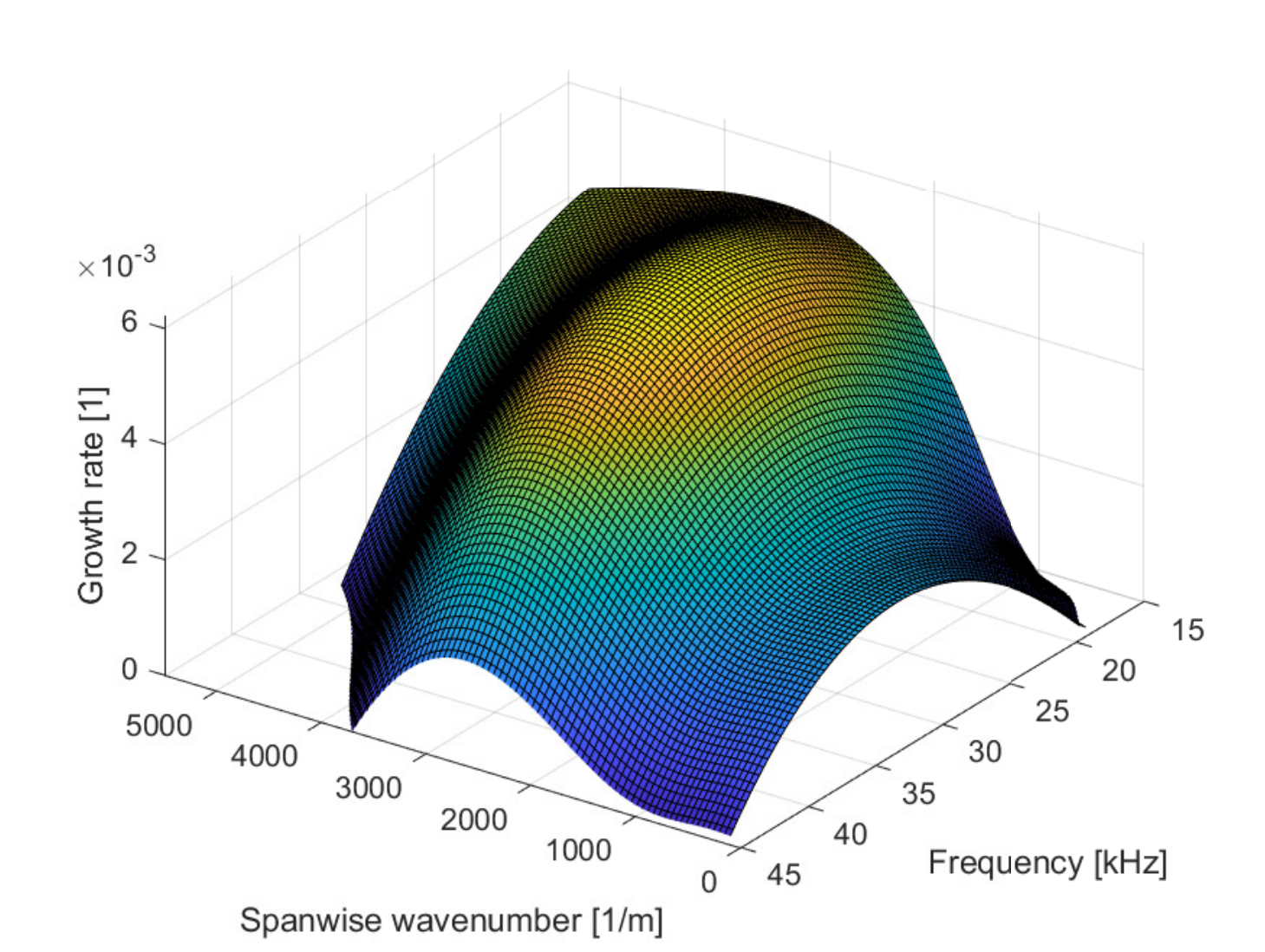}
        \caption{$x/L=50\%$}
        \label{fig_CRM_3D_scan_x05}
    \end{subfigure}
    \hfill
    \begin{subfigure}[b]{0.33\textwidth}
        \centering
        \includegraphics[width=\textwidth]{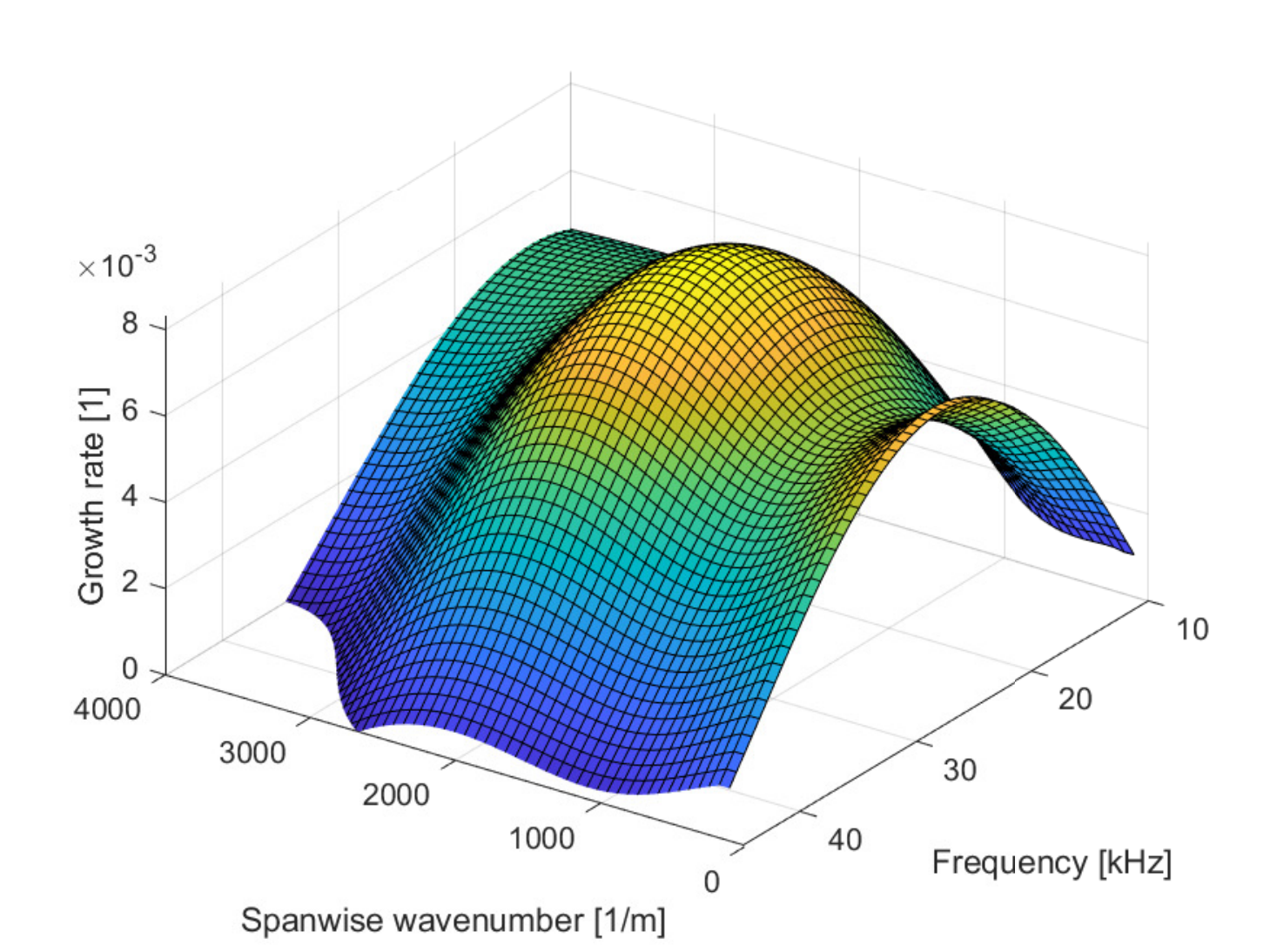}
        \caption{$x/L=70\%$}
        \label{fig_CRM_3D_scan_x07}
    \end{subfigure}
    \caption{Spanwise wavenumber-frequency scan using 3D baseflow profile.}
    \label{fig_CRM_3D_scan}
\end{figure}

\begin{figure}[hbt!]
    \centering
    \begin{subfigure}[b]{0.33\textwidth}
        \centering
        \includegraphics[width=\textwidth]{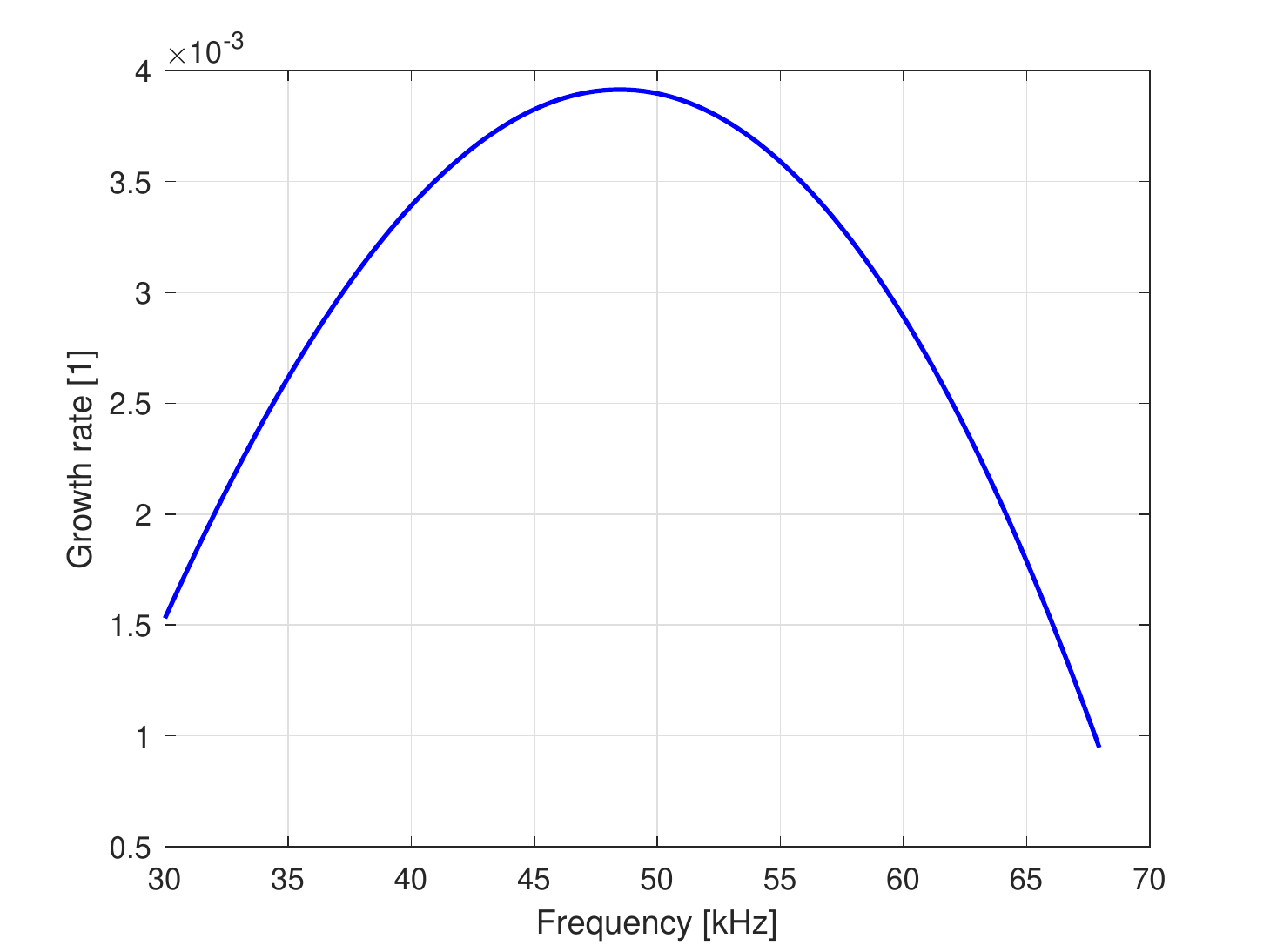}
        \caption{$x/L=30\%$}
        \label{fig_CRM_2D_scan_x03}
    \end{subfigure}
    \hfill
    \begin{subfigure}[b]{0.33\textwidth}
        \centering
        \includegraphics[width=\textwidth]{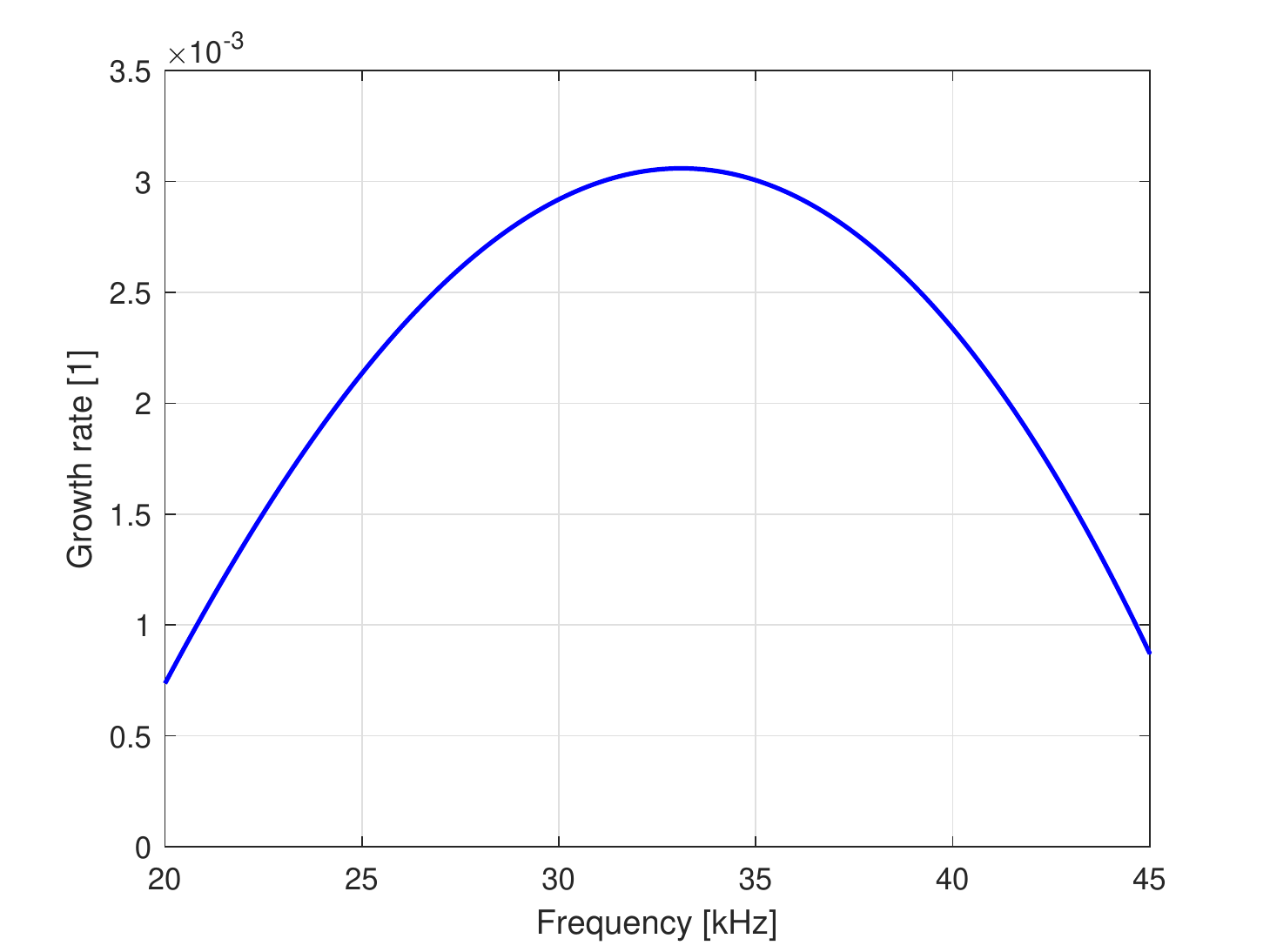}
        \caption{$x/L=50\%$}
        \label{fig_CRM_2D_scan_x05}
    \end{subfigure}
    \hfill
    \begin{subfigure}[b]{0.33\textwidth}
        \centering
        \includegraphics[width=\textwidth]{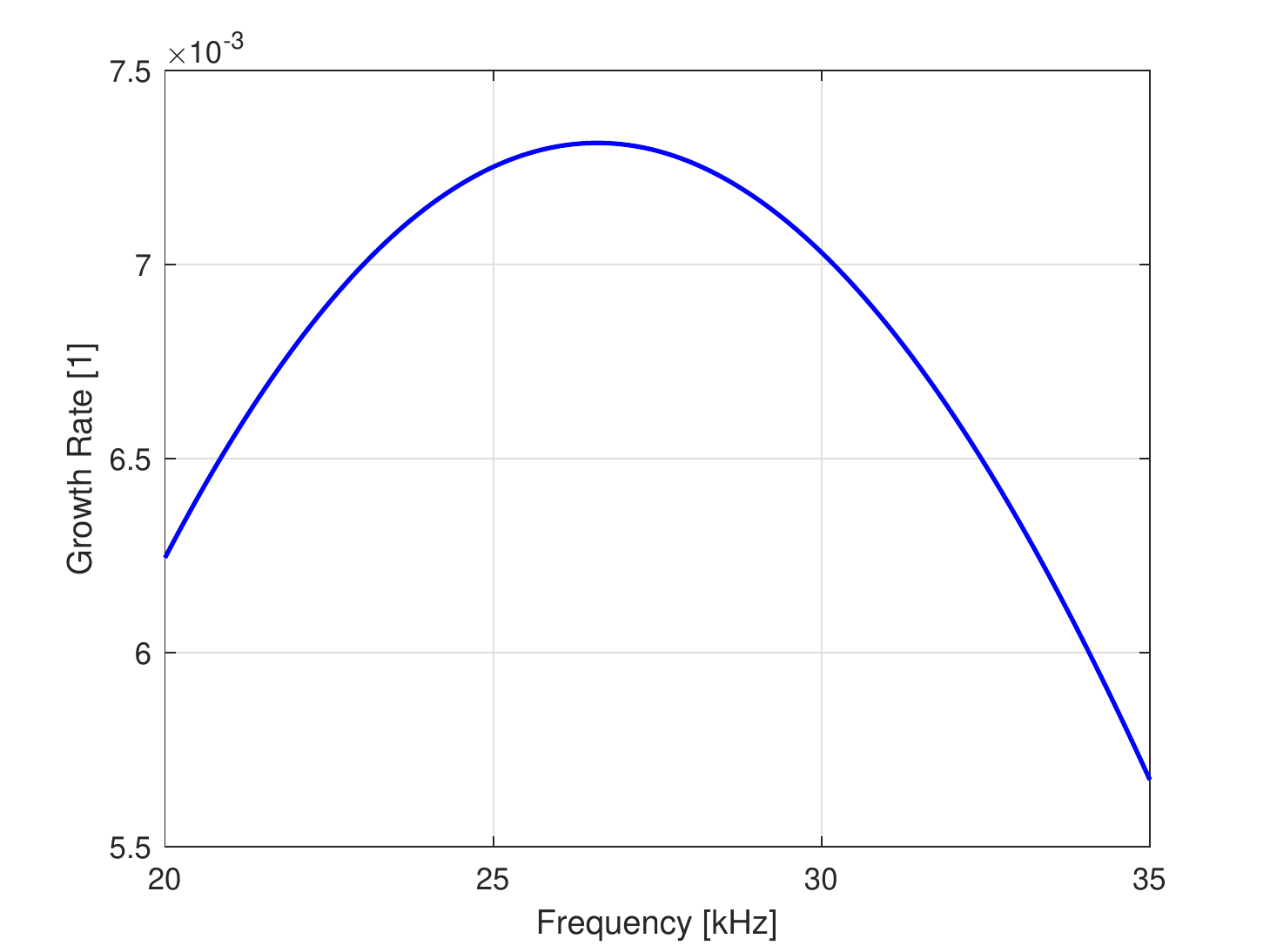}
        \caption{$x/L=70\%$}
        \label{fig_CRM_2D_scan_x07}
    \end{subfigure}
    \caption{Frequency scan using 2D baseflow profile.}
    \label{fig_CRM_2D_scan}
\end{figure}

\begin{figure}[hbt!]
  \centering
  \includegraphics[width=0.49\textwidth]{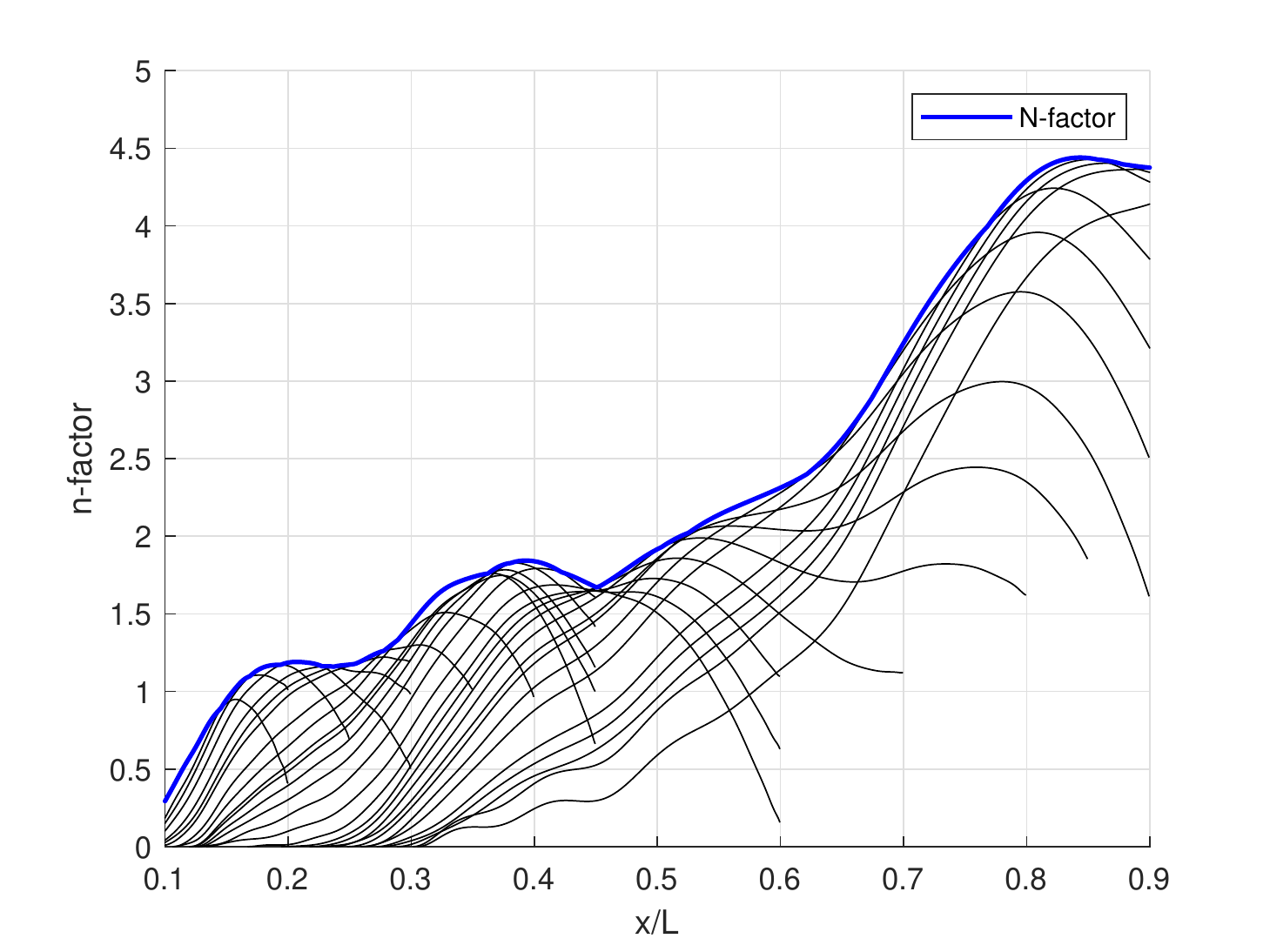}
  \caption{$N$-factor for the 2D wing section of CRM-NLF geometry.}
  \label{fig_CRM_2D_Nfactor}
\end{figure}

Fig. \ref{fig_CRM_3D_scan} shows the growth rate of the TS waves in the parametric space at $x/L=30\%$, $x/L=50\%$, and $x/L=70\%$. The most amplified TS waves have a spanwise wavenumber range of $[2000, 4000]$ 1/m and the frequency range of $[27,50]$ kHz. In the result it is noted that the growth rate has a similar variation tendency with frequency at low spanwise wavenumber and in the most amplified range. We therefore carried out a frequency scan at the same streamwise position using the 2D baseflow profiles. As is shown in Fig. \ref{fig_CRM_2D_scan}, the frequency of the most amplified TS waves in 2D baseflow also approximately sits in the range $[27,50]$ kHz. Since the 2D baseflow has a similar frequency sensitivity to the 3D baseflow, as a preliminary analysis we simulate the TS wave development in the 2D baseflow, and the the amplification curves are given in Fig. \ref{fig_CRM_2D_Nfactor}, where each $n$-factor is computed based on disturbance on the streamwise velocity in the body-fitted coordinate system. The variation of the $N$-factor curve is not as monotonic as that for the clean flat plate case in Fig. \ref{fig_FP_clean_N} since the pressure gradient on the upper surface keeps shifting between favourable and adverse while the TS waves are more amplified by the adverse type. Moreover, the TS waves can be suppressed by the favourable gradient, which causes the drop between $x/L=82\%$--$90\%$.

%%%%%%%%%%%%%%%%%%%%%%%%%%%%%%%%%%%%%%%%%%%%%%%%%%%%%%%%%%%%%%%%%%%%
\section{Conclusion}
\label{sec::condlusion}
% A conclusion might elaborate on the importance of the work or suggest applications and extensions.

The adoption of the $e^N$ method for transition analysis involves two major steps as the computation for baseflow and disturbance field. Between these two steps the linear stability analysis is preferred to target the parametric space of interest for the disturbance frequency and spanwise wavenumber, with which the linear growth of certain disturbances dominate the computational domain and finally support the $N$-factor curve. The steps in the above workflow involves different kinds of simulations, and different computational tools are therefore used particularly when the geometries are complex. The resultant data conversion issue causes not only precision losses of data but also extra workload for researchers and engineers. In the current work an open-source and unified framework for laminar boundary layer natural transition analysis is introduced. We extended the applicability to transonic compressible flows over complex geometries. The major challenges in the workflow are discussed and solutions are provided. 

For the first major step, we compute the baseflow in a near wall, reduced domain to lower the computational cost. The computation in the reduced domain relies the boundary data interpolated from a pre-generated RANS simulation, where the pressure distribution is considered well predicted as examined by many experiments. A successful baseflow therefore needs to have pressure compatibility with the background RANS result. However, because of the hyperbolic nature for compressible flows, the pressure compatibility is not automatically guaranteed in the standard Riemann inflow boundary condition enforcement, and significant deviation can be caused for complex geometries. In addition, there are more than one type of pressure compatible inflow whereas not all of them are numerically stable. To figure out the best practice for the pressure compatible inflow, we revisit the 1D stability analysis for DG based simulation, and construct entropy-pressure, velocity-pressure, and momentum-pressure compatible inflows. It is found that only the entropy-pressure compatible one  is stable. Additional support for the entropy-pressure compatibility includes the compatibility on the speed of sound and density, and tangential momentum components in multi-dimensional simulations.

In the second major step to predict the growth of disturbances, data contamination caused by wave reflection is the main problem. The wave reflection stems from the adoption of blowing-suction to excite the disturbances developing in the boundary layer and the reflective inflow boundary condition for the truncated computational domain, particularly when the domain is reduced. A sponge region can suppress the reflected waves to a desired level (2 or more order smaller than the boundary layer disturbance of interest). Another source for wave reflection is the potential interaction between disturbances and surface irregularities on the wall, such as a small step embedded in the boundary layer. In this case the reflected waves would have the same frequency as the incident wave and therefore the signal separation is difficult. Since limited analysis have been reported, signal separation through wave modelling is not applicable at the moment. We choose to reduce the contamination through filtering as a more user-friendly approach. However, wave reflection features at surface irregularities is under further study, which may bring us better solution in the future.

The workflow is then verified through a Mach $0.8$ flat plate case with both clean and stepped geometries. The generated $N$-factors for the TS waves well agree with the reference result. The applicability of the framework is evaluated through the precision study and the linear growth limit of TS waves are reported. Finally, the transitional performance over a wing section of the CRM-NLF model is studied going through the full workflow. The $N$-factor is generated on the 2D baseflow as a preliminary result.

\section*{Acknowledgments}
The authors acknowledge support from the Beijing Aircraft Technology Research Institute of COMAC from 2019 to 2021.

\bibliography{sample}

\end{document}